\documentclass{emulateapj}
\usepackage{times,common}

\begin{document}

\markboth{C.-J. Ma et al.}{U/LIRGs in XCS\,J2215 at $z=1.46$}
\title{Dusty starbursts and the formation of elliptical galaxies:\\ A SCUBA-2 survey of a \textit{z}\,=\,1.46 cluster\footnote{{\it Herschel} is an ESA space observatory with science instruments provided by European-led Principal Investigator consortia and with important participation from NASA.}}
\author{
C.-J.\ Ma\altaffilmark{2},
Ian Smail\altaffilmark{2,3}, 
A.\,M.\ Swinbank\altaffilmark{3}, 
J.\,M.\ Simpson\altaffilmark{2},
A.\,P.\ Thomson\altaffilmark{2},
C.-C.\ Chen\altaffilmark{2},\\
A.\,L.\,R.\ Danielson\altaffilmark{3},
M.\ Hilton\altaffilmark{4}, 
K.\ Tadaki\altaffilmark{5},
J.\,P.\ Stott\altaffilmark{3} \&
T.\ Kodama\altaffilmark{5,6}
}
\altaffiltext{2}{Centre for Extragalactic Astronomy, Department of Physics, Durham University, South Road, Durham DH1 3LE, UK}
\altaffiltext{3}{Institute for Computational Cosmology, Durham University, South Road, Durham DH1 3LE, UK}
\altaffiltext{4}{Astrophysics \& Cosmology Research Unit, School of Mathematics, Statistics \& Computer Science, University of KwaZulu-Natal, Private Bag X54001, Durban 4000, South Africa}
\altaffiltext{5}{National Astronomical Observatory of Japan, Mitaka, Tokyo 181-8588, Japan}
\altaffiltext{6}{Department of Astronomical Science, The Graduate University for Advanced Studies, Mitaka, Tokyo 181-8588, Japan}

\begin{abstract}
We report the results of a deep SCUBA-2 850- and 450-$\mu$m survey for
dust-obscured ultra-/luminous infrared galaxies (U/LIRGs) in the field of the $z$\,=\,1.46 cluster
XCS\,J2215.9$-$1738.  We detect a striking overdensity of sub-millimeter
sources coincident with the core of this cluster: $\sim$\,3--4\,$\times$ higher than
expected in a blank field.  We use the likely radio
and mid-infrared counterparts to show that the bulk of these sub-millimeter sources have spectroscopic or
photometric redshifts which place them in the cluster and that their
multi-wavelength properties are consistent with this association.  The
average far-infrared luminosities of these galaxies are
(1.0\,$\pm$\,0.1)\,$\times$\,10$^{12}$\,L$_\odot$, placing them on the U/LIRG boundary.  Using the total
star formation occurring in the obscured U/LIRG population within the cluster we show that the resulting
mass-normalized star-formation rate for this system supports previous claims of a rapid
increase in star-formation activity in cluster cores out to
$z\sim$\,1.5, which must be associated with the on-going formation of
the early-type galaxies which reside in massive clusters today.
\end{abstract}


\section{Introduction}\label{sec:intro}  

The relationship between star formation and local galaxy density  is one of the clearest pieces of
evidence of the effect of environment on galaxy evolution.  In the
local Universe the star-formation rate (SFR) of galaxies increases
with galaxy density up to the scale of galaxy groups, and then drops
sharply in the denser environment of clusters
\citep[e.g.][]{lewis02,gomez03}.  As a consequence, galaxy clusters are
dominated by passive early-types (ellipticals and lenticular
galaxies) whose stellar populations
indicate little recent star-formation activity, while the low-density field is populated by
star-forming spiral galaxies \citep[e.g.][]{dressler97,smith05}.  As all galaxies must
have been ``star forming'' at some time, this difference suggests that some
feature of dense environments has acted to transform star-forming
galaxies into the passive populations which dominate these regions
today.

However, the fraction of star-forming galaxies (or equally late-type, disk
galaxies) in clusters increases rapidly with redshift
\citep[e.g.][]{dressler97}, which suggests that the environment of
growing clusters may have been less hostile in the past.  Indeed,
 mid-infrared and sub-millimeter wavelength observations have
 revealed populations of strongly star-forming (ultra-)luminous
infrared galaxies (U/LIRGs) in more distant clusters
\citep[e.g.][]{brodwin13,pintoscastro13,smail14,alberts14,santos14, santos15}.  This
enhanced activity means that the mass-normalized integrated star-formation rate for massive clusters evolves rapidly with redshift,
$\Sigma({\rm SFR}) / {\rm M}_{\rm cl} \propto (1+z)^{\alpha}$ with $\alpha$ as high as seven
\citep[e.g.][]{geach06,popesso11,popesso12}.  Below $z\sim$\,1, these vigorously
star-forming systems are generally found on the outskirts of clusters
and are inferred to be in-falling, while the star-formation activity
is still suppressed at the  centers of clusters
\citep[e.g.][]{santos13}.  However, at $z\sim$\,1.5 and beyond,
strongly star-forming galaxies have been identified in the cluster
cores \citep{hayashi10,tadaki12}. This has been claimed to represent
the ``reversal'' of the star formation--density relation
\citep[e.g.][]{tran10,santos15}.
%
%fig 1
%
\begin{figure*}[th]
\epsscale{1.0}\plottwo{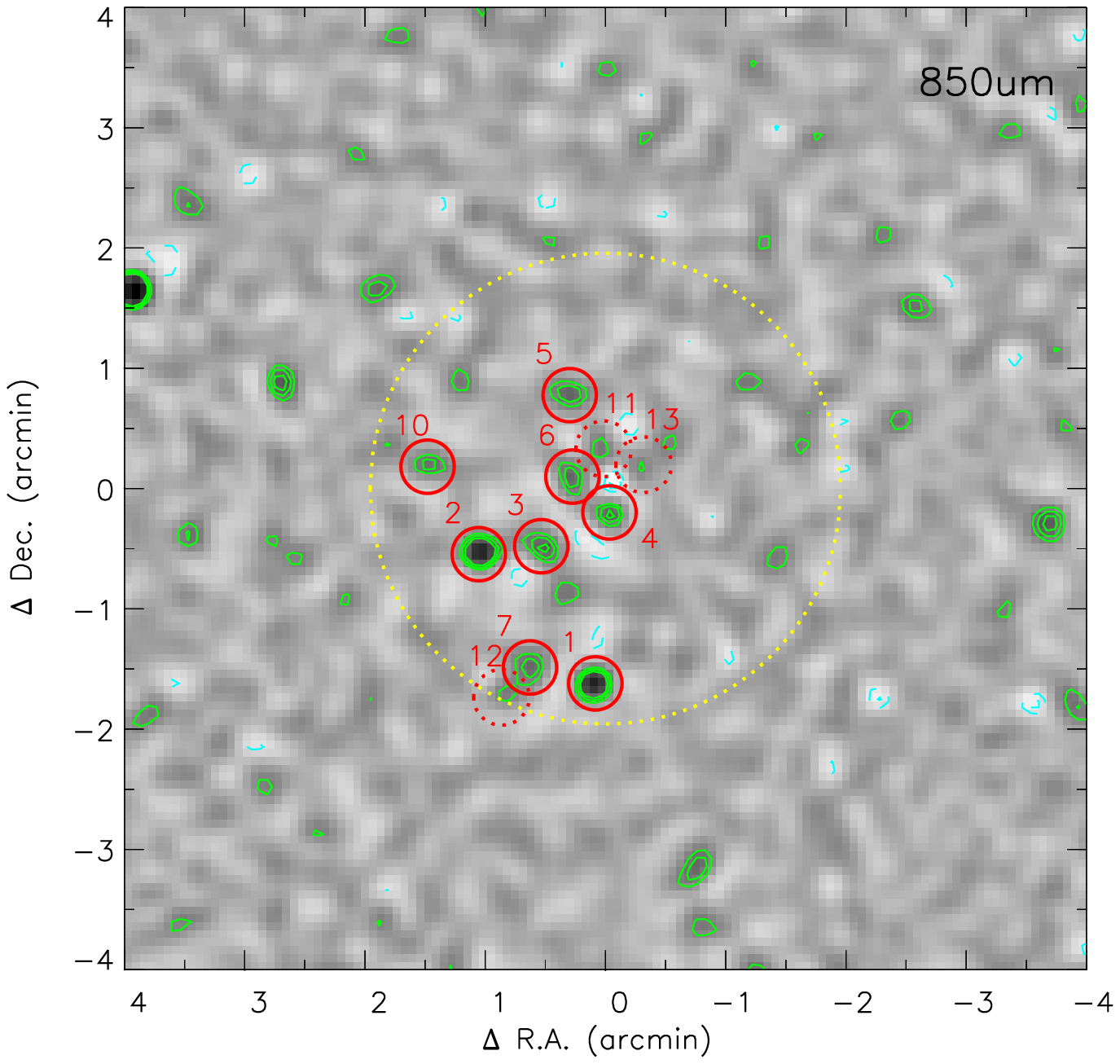}{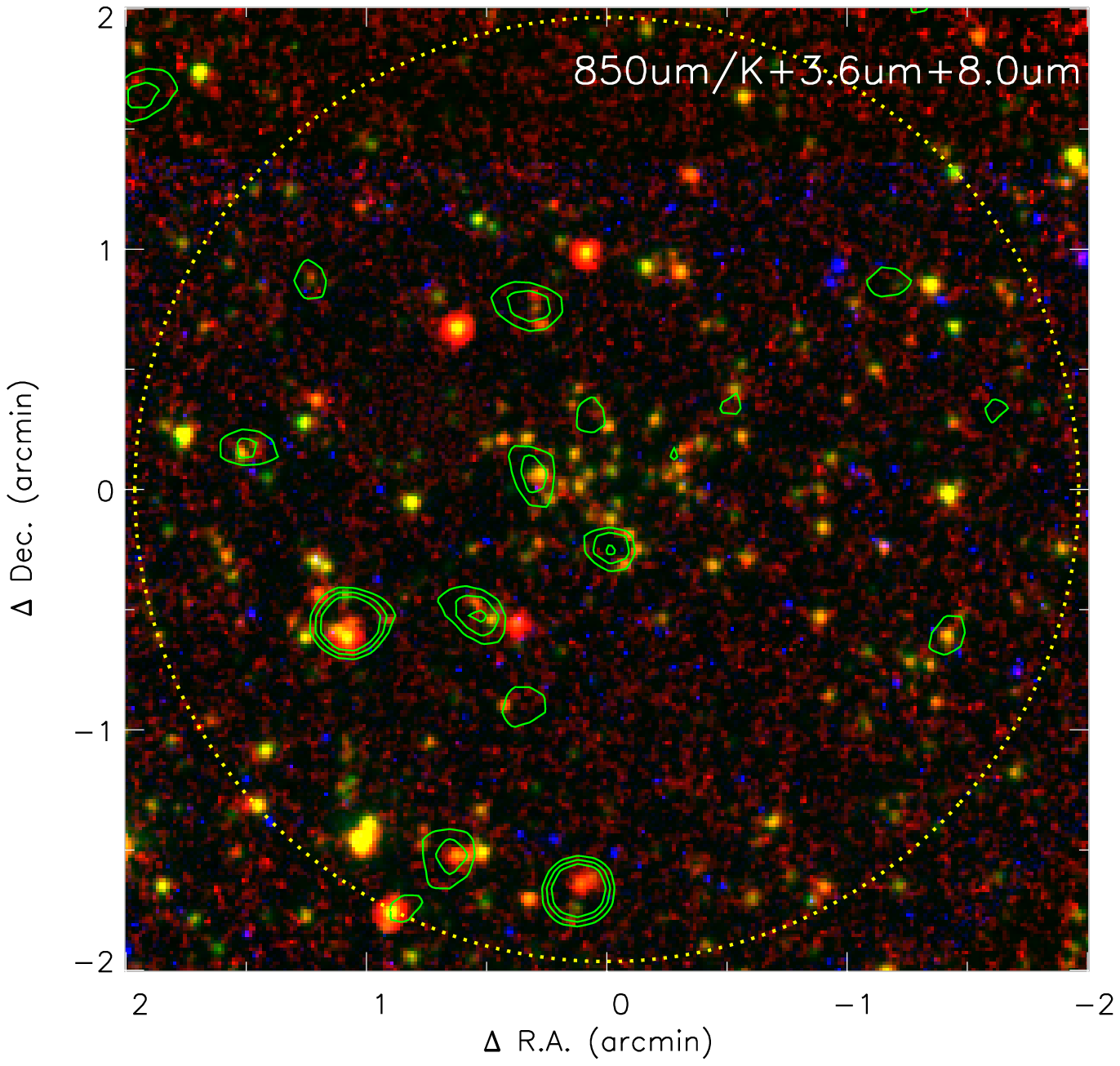}
\figcaption{{\it Left}: An 8$\arcmin\times $\,8$\arcmin$ area around
  the core XCS\,J2215 from our 850\,$\mu$m SCUBA-2 map with the detected
  sources within a 1\,Mpc projected radius of the cluster center
  marked: the red solid circles indicate the 4-$\sigma$ detections at
  850\,$\mu$m, while the red dashed circles mark the fainter
  850\,$\mu$m sources ($>$\,3$\sigma$) which are also simultaneously
  detected in all of the {\it Herschel} PACS 70\,$\mu$m, 160\,$\mu$m,
  and MIPS 24\,$\mu$m bands.  We see a strong over-density of
  sub-millimeter galaxies (SMGs) in the core regions of the cluster,
  which we associate with a population of ULIRG cluster
  members.  The large dashed yellow circle shows the 1\,Mpc radius and
  we also plot the 3, 4, 5\,$\sigma$ positive (green) and negative
  (cyan) contours of the 850\,$\mu$m map.  {\it Right}: A true-color image of the central
  1-Mpc radius region highlighted on the left-hand panel (again shown
  by the large dashed yellow circle).  This is constructed from $K_s$,
  3.6\,$\mu$m and 8\,$\mu$m images and we overlay the same SCUBA-2
  850\,$\mu$m signal-to-noise contours. Several of the SMGs coincide
  with galaxies with very red near-infrared colors, much redder than
  the bulk of the passive cluster galaxies which define the strong overdense
  population in the cluster core.
 }\label{fig:s850} 
\end{figure*}

XMMXCS\,J2215.9$-$1738 (XCS\,J2215 hereafter) at $z$\,=\,1.46 is an
excellent target to study the nature of star-formation activity in a
high-redshift cluster.  It is one of the most distant cluster
discovered in X-rays \citep{stanford06}, demonstrating that it
represents a deep gravitational potential.  A series of extensive
multi-wavelength follow-up programmes
\citep[e.g.][]{hilton07,hilton09,hilton10} confirm that XCS\,J2215 is
indeed a relatively well-developed cluster.  Using {\it XMM-Newton}
and {\it Chandra} observations \citet{hilton10} measured an X-ray
luminosity of $L_{\rm
  X}$\,=\,2.9$^{+0.2}_{-0.4}\times$\,10$^{44}$\,erg\,s$^{-1}$ and a
temperature of $T_{\rm X}$\,=\,4.1$^{+0.6}_{-0.9}$\,keV.  The velocity
dispersion for galaxies within the virial radius
($R_{200}$\,=\,0.8\,Mpc) is $\sigma_{\rm
  v}$\,=\,720\,$\pm$\,110\,\,km\,s$^{-1}$, suggesting that the cluster
lies on the self-similar $\sigma_{\rm v}$--$T_{\rm X}$ scaling
relation \citet{hilton10}.  This then implies a cluster mass of
$M_{\rm cl}\sim $\,3$\times$\,10$^{14} $\,M$_{\odot}$ using the
$M_{\rm cl}$--$\sigma_{\rm v}$ relation \citep[e.g.][]{koyama10}.
However, the cluster is unlikely to be virialized since it lies below
the local $L_{\rm X}$--$T_{\rm X}$ relation
\citep[e.g][]{markevitch98}, and there is also mild bimodality in the
velocity distribution of the 44 cluster members identified in
\citet{hilton10}.

XCS\,J2215 was the one of the first clusters in which a reversal in
the star formation--density relation  was
identified. \citet{hayashi10} found twelve [O{\sc ii}] emitters
within 0.25\,Mpc of the cluster center \citep[see
  also][]{hayashi11,hayashi14}.  In addition, \citet{hilton10} found
two mid-infrared bright, apparently star-forming, cluster members
within 0.2\,Mpc of the center using {\it Spitzer}\,/\,MIPS.  These
studies suggest that both high- and low-SFR galaxies
are found in the densest regions of this cluster, and they showed that
the star-formation density does not drop at the cluster core.
However, at the redshift of XCS\,J2215 estimating the total
SFR from the {\it Spitzer} MIPS 24\,$\mu$m luminosity is complicated by the
presence of both Polycyclic Aromatic Hydrocarbon (PAH) emission and
potentially strong Silicate absorption features in the band, as well
as potential contamination from continuum emission due to AGN.
Similarly, as we demonstrate later in the paper, estimating the total SFRs from [O{\sc ii}] fluxes also suffers
from uncertainties due to the dust obscuration.  An independent route
to identify strongly star-forming galaxies and determine their SFR is to use the dust thermal emission in the far-infrared
and sub-millimeter wavebands.  In this paper, we have therefore
combined new SCUBA-2 sub-millimeter observations of XCS\,J2215 with
archival mid- and far-infrared and radio data to study the SFR and dust properties of the galaxy population in this
$z$\,=\,1.46 cluster.

Throughout this paper, we adopt a $\Lambda$CDM cosmology with
$H_0=$\,70\,km\,s$^{-1}$\,Mpc$^{-1}$, $\Omega_{\Lambda}=$\,0.7, and $\Omega_m =$\,0.3.  In this
cosmology, 1$\arcsec$ corresponds to 8.5\,kpc at the cluster redshift. The cluster center derived from the X-ray centroid is at 22\,15\,59.5, $-$17\,38\,03 (J2000) \citep{hilton10}. Magnitudes are quoted in the AB system.

\section{Observation and Data Reduction} \label{sec:data}

This work is primarily based on the sub-millimeter maps at 450- and 850-$\mu$m
 obtained with the Sub-millimeter Common-User Bolometer Array 2
\citep[SCUBA-2,][]{holland13} on the James Clerk Maxwell Telescope
(JCMT).  In addition, we compile multi-wavelength images and
photometry of the sub-millimeter galaxies from various archives,
including {\it Spitzer}, {\it Herschel}\footnote{{\it Herschel} is an ESA
  space observatory with science instruments provided by European-led
  Principal Investigator consortia and with important participation
  from NASA.}, {\it Hubble Space Telescope} and the Karl G.\ Jansky
Very Large Array (JVLA).  We also collate redshifts and optical/near-infrared data for galaxies in
this field from \citet{hilton10} \citep[see also][]{hilton07,hilton09,hayashi10}.

\subsection{SCUBA-2 Observations} \label{sec:data:SMG}

The SCUBA-2 observations of XCS\,J2215 are performed in 2013
July--August in Band 1 conditions ($\tau_{225}\leq $\,0.05) as part of projects M13AU29 and
M13BU10.  The simultaneous 450 and 850\,$\mu$m observations yield a
total integration of 8\,hrs, as twelve 40\,min scans, centered on the cluster.  The data from each scan are processed
individually to make maps using the {\sc smurf}\footnote{http://starlink.jach.hawaii.edu/docs/sc21.htx/sc21.html} reduction software
package \citep{jenness13,chapin13}.  These maps, in unit of pW, are calibrated to Jy using the
canonical calibration factor of
FCF$_{450}$\,=\,491\,$\pm$\,67\,Jy\,beam$^{-1}$\,pW$^{-1}$ and
FCF$_{850}$\,=\,537\,$\pm$\,24\,Jy\,beam$^{-1}$\,pW$^{-1}$ at 450 and
850\,$\mu$m respectively, which are consistent with the calibration from flux standards obtained on the relevant nights.  We coadd the individual exposures using the {\sc picard} package \citep[e.g.][]{jenness08} to construct the final map in each
band.  Finally, we apply a matched-filter to the maps with beam
full-width-half-maximum of 8$\arcsec$ and 15$\arcsec$ at 450\,$\mu$m
and 850\,$\mu$m respectively to improve the sensitivity for point
source detection.  The 1-$\sigma$ noise levels at the center of the
final maps are 5.4 and 0.63\,mJy\,beam$^{-1}$ at 450 and 850\,$\mu$m
respectively. The sensitivity drops to 50\% at the radius of $\sim$\,5.4$\arcmin$ from the map center due to the scan coverage of the {\sc daisy} pattern.
The central 8$\arcmin\times$\,8$\arcmin$ area of the
match-filtered 850\,$\mu$m image is shown in the Fig.~\ref{fig:s850}.

To determine the accuracy of the absolute astrometry of the
850\,$\mu$m map, we compare the coordinates of 850\,$\mu$m sources
(detected at $>$\,3$\sigma$) with the closest matched counterparts at
24\,$\mu$m.  After removing three outliers, the mean
offset of the 17 counterparts at 24\,$\mu$m is $\Delta{\rm
  RA}=$\,1.4$\arcsec\pm$\,0.6$\arcsec$ and $\Delta{\rm
  Dec}=-$0.4$\arcsec\pm$\,0.8$\arcsec$. We correct the astrometry of the
850\,$\mu$m map using the 24\,$\mu$m offsets. 
We similarly confirm that astrometry of the 450\,$\mu$m  map is consistent using the counterparts of
sources at 24\,$\mu$m.   The mean offsets are $\Delta{\rm RA}=$\,-0.2$\arcsec\pm$\,0.6$\arcsec$ and $\Delta{\rm  Dec}=$\,0.1$\arcsec\pm$\,0.6$\arcsec$. Thus, there is no significant offset identified and so we choose not to correct the astrometry.

%
%fig 2
%
\begin{figure*}
\includegraphics[width=0.15\textwidth]{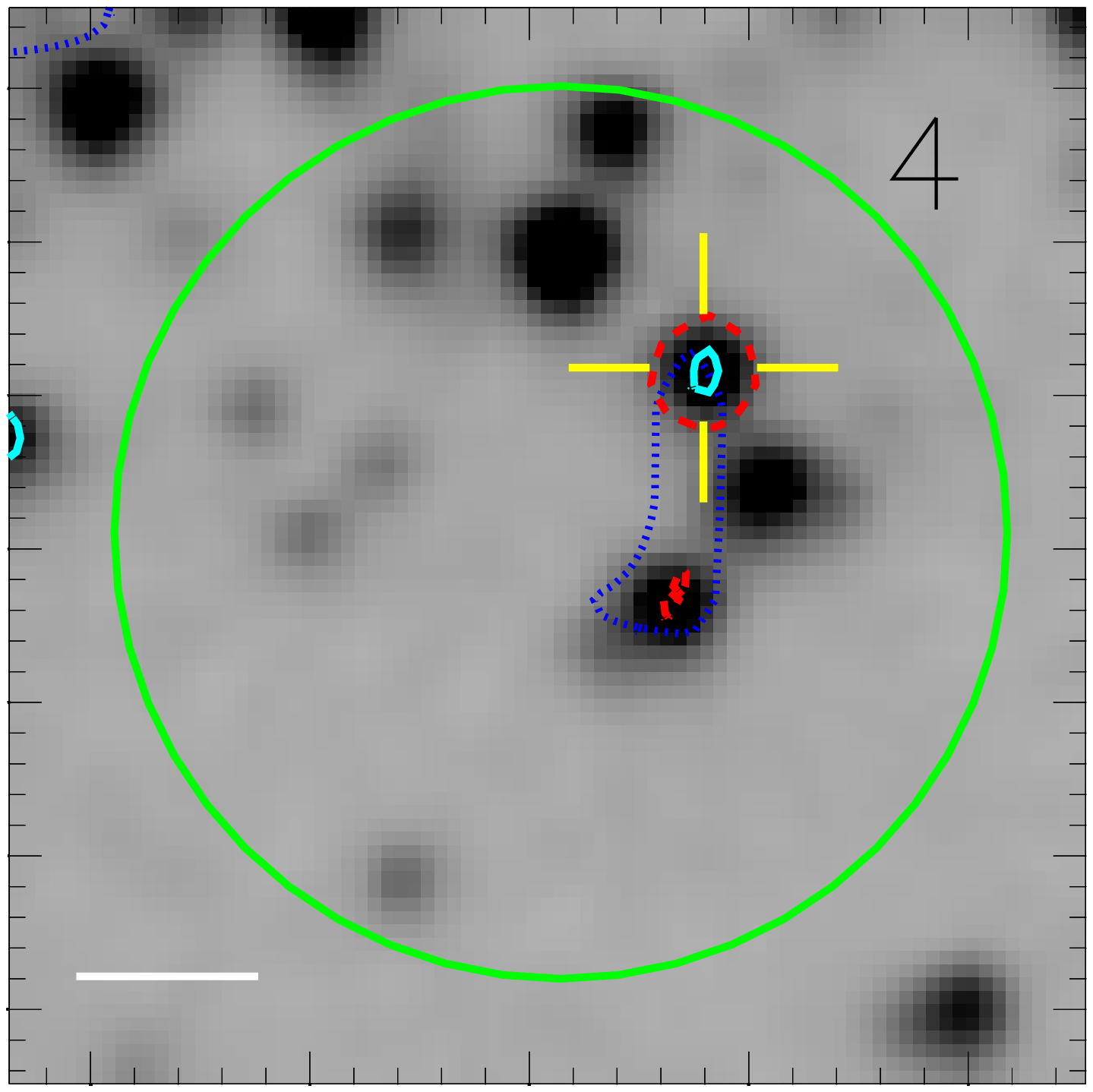} \includegraphics[width=0.15\textwidth]{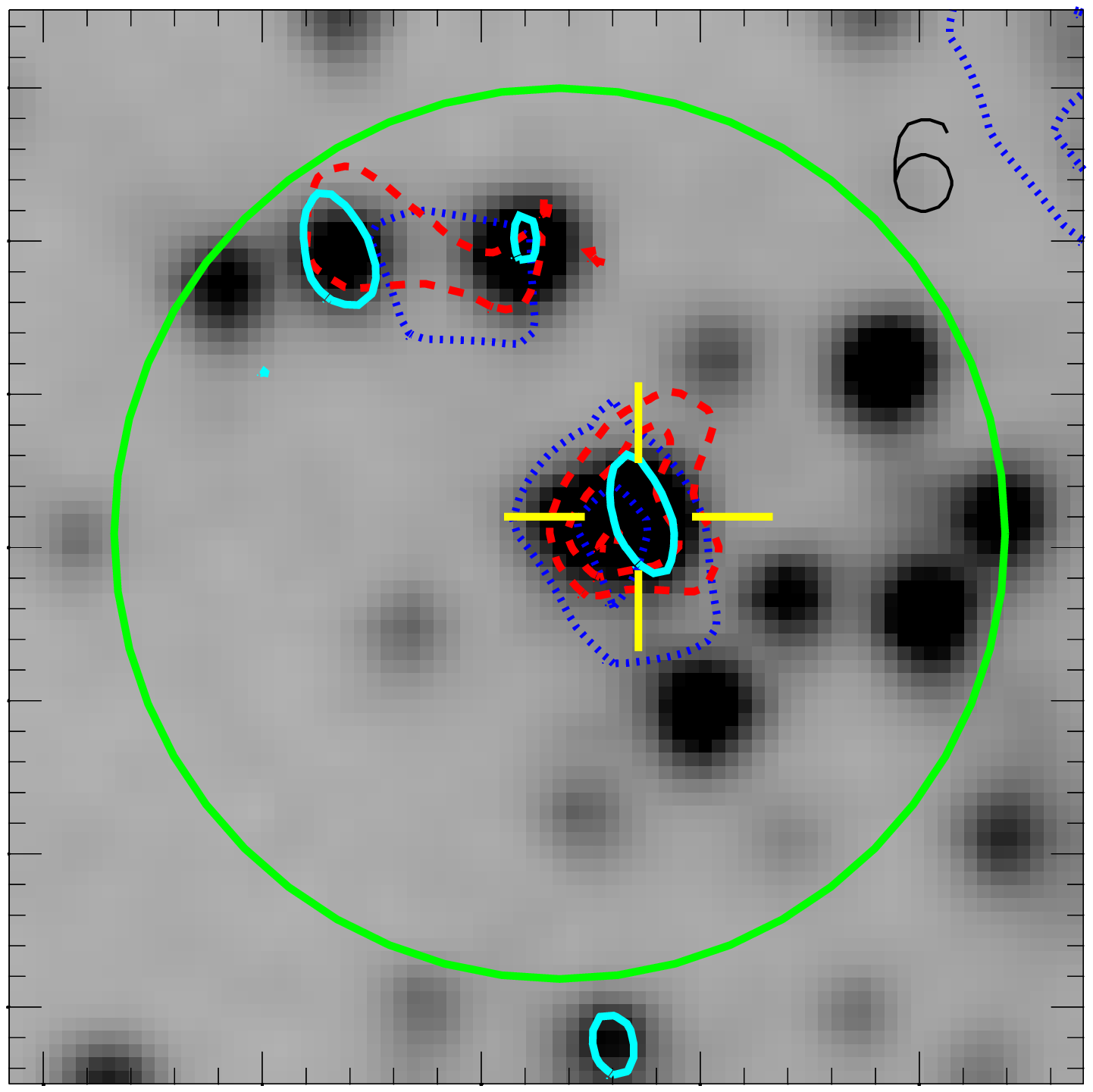}\hspace*{0.02\textwidth}
\includegraphics[width=0.15\textwidth]{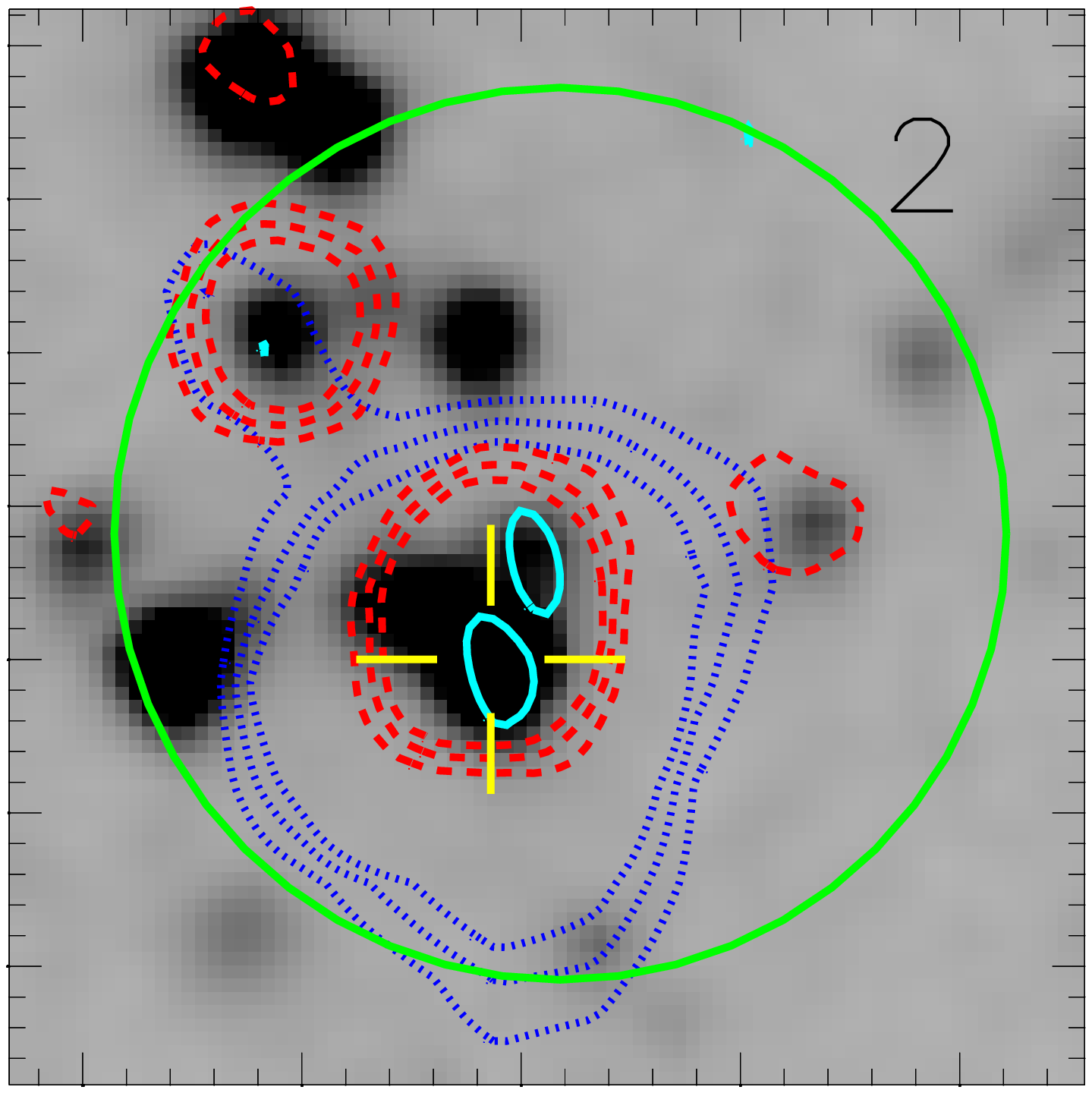}\includegraphics[width=0.15\textwidth]{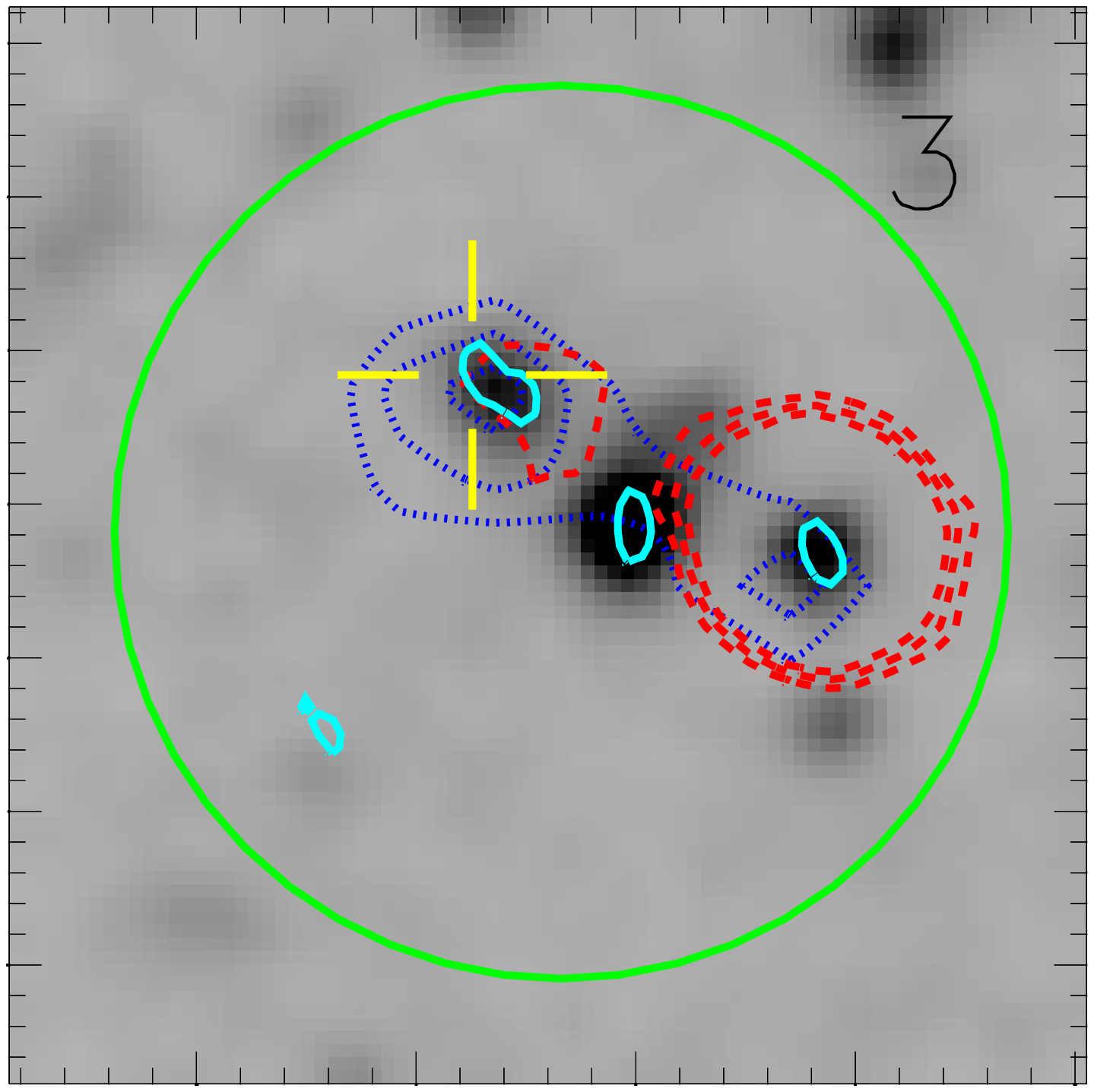} \hspace*{0.02\textwidth} \includegraphics[width=0.15\textwidth]{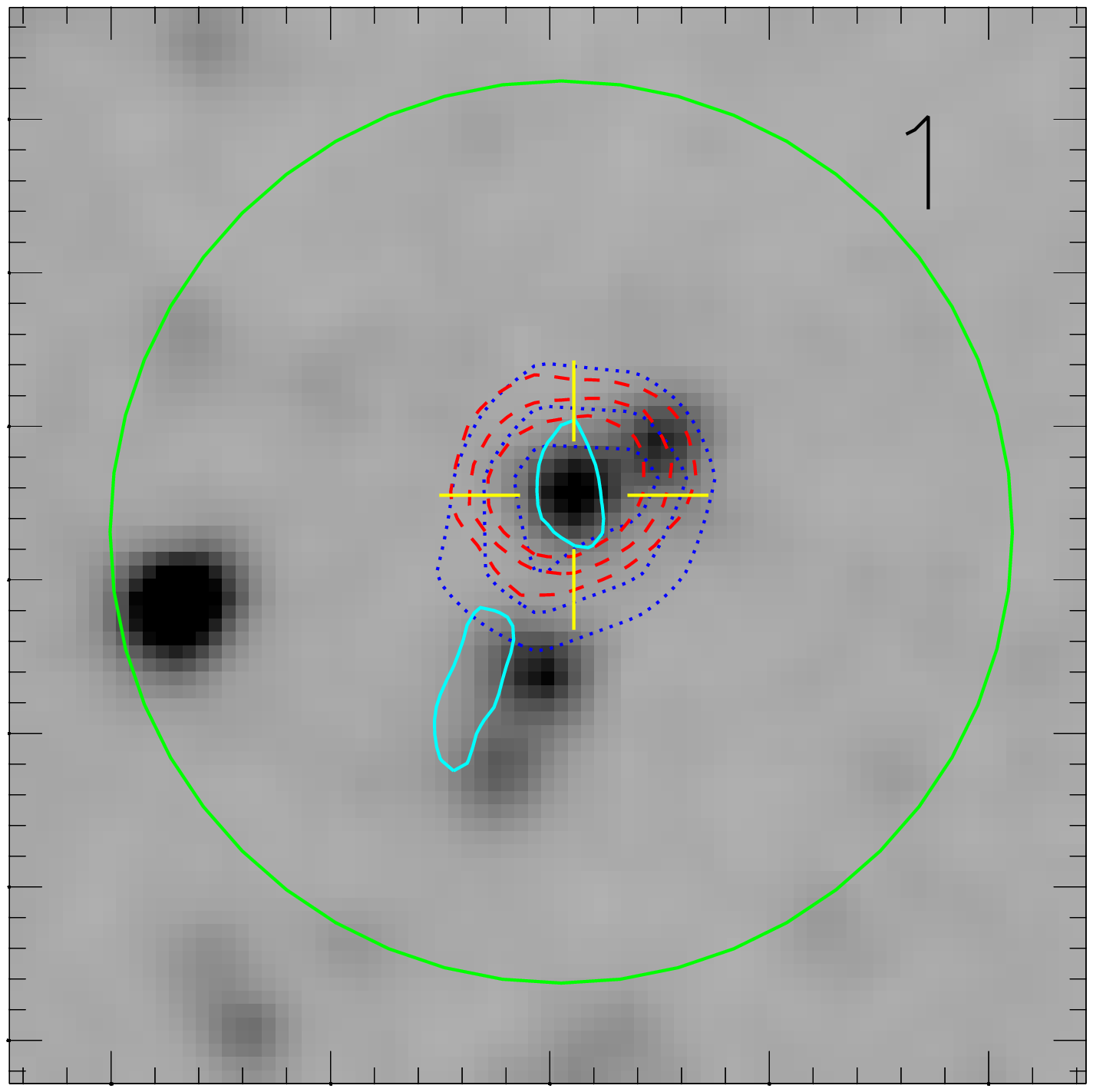} \includegraphics[width=0.15\textwidth]{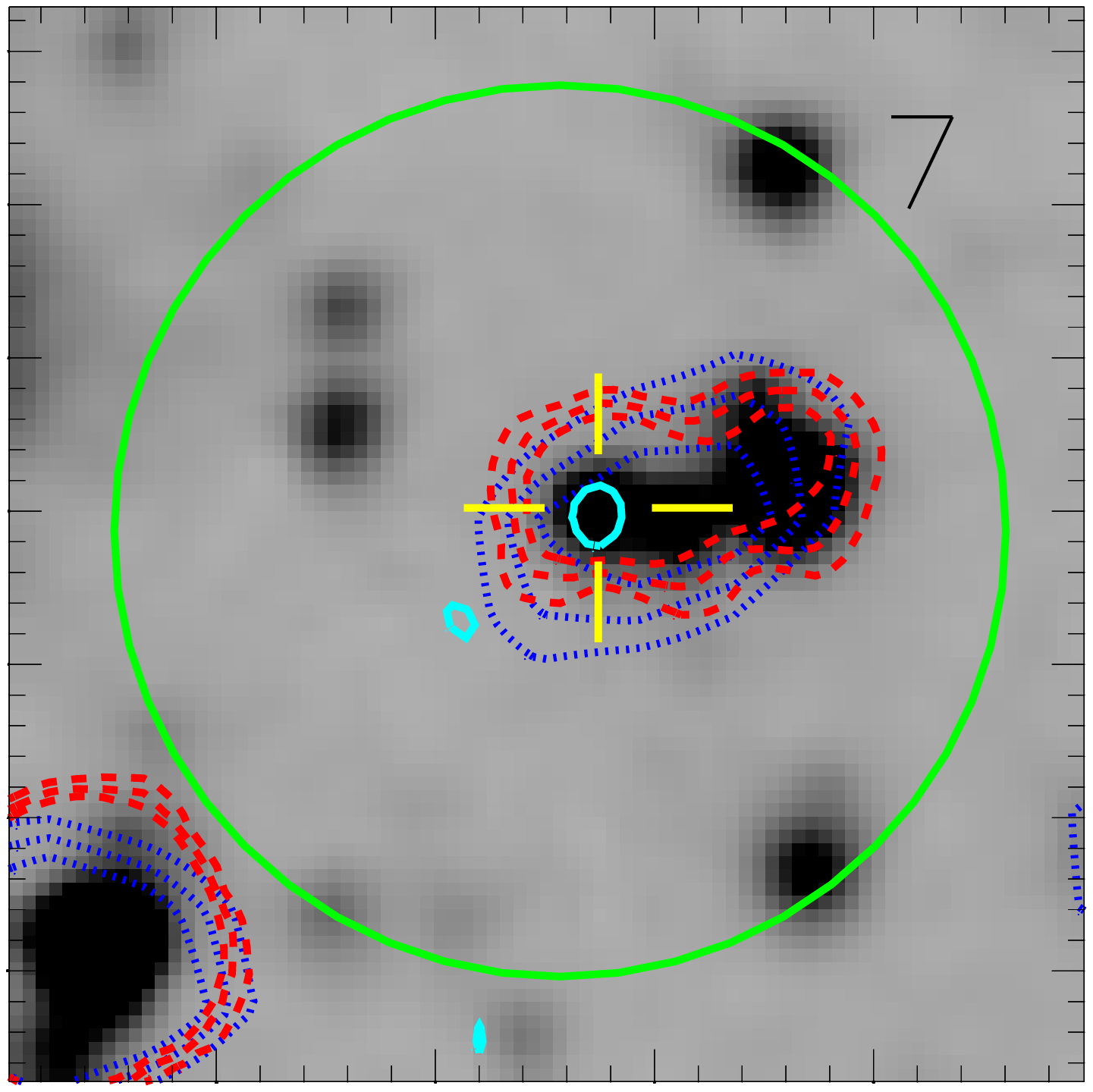}\\
\includegraphics[width=0.15\textwidth]{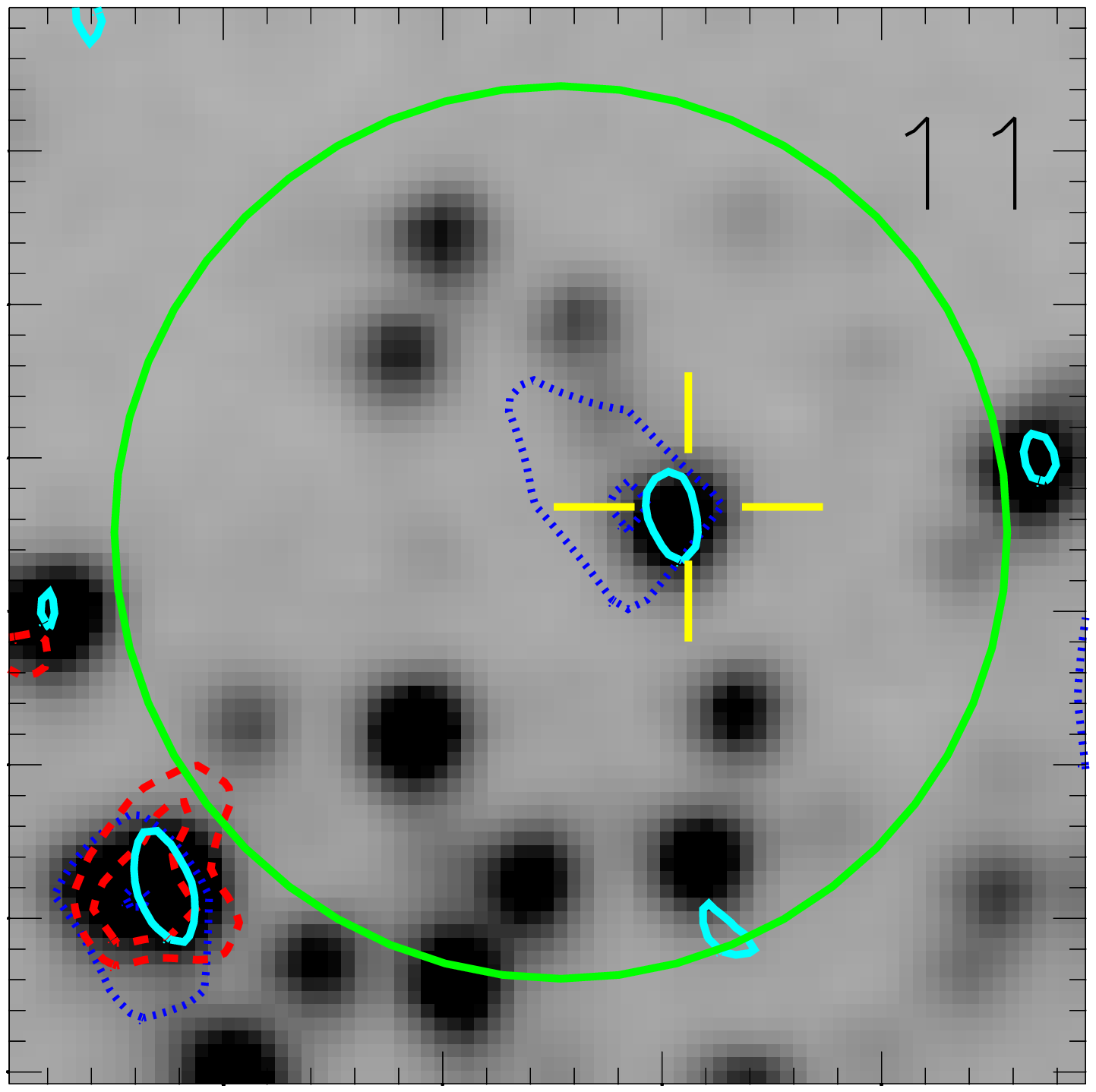} \includegraphics[width=0.15\textwidth]{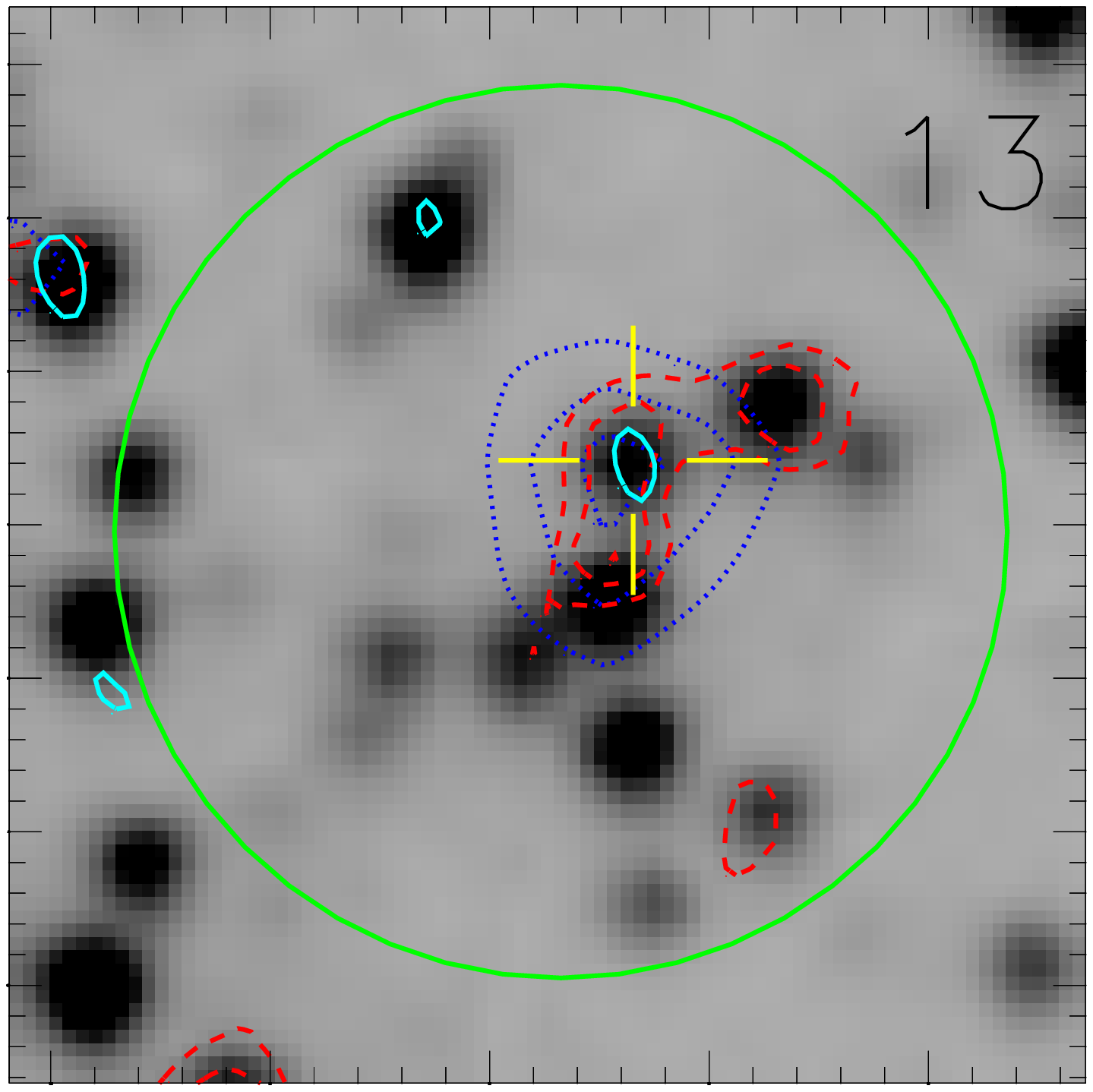}\hspace*{0.02\textwidth}
\includegraphics[width=0.15\textwidth]{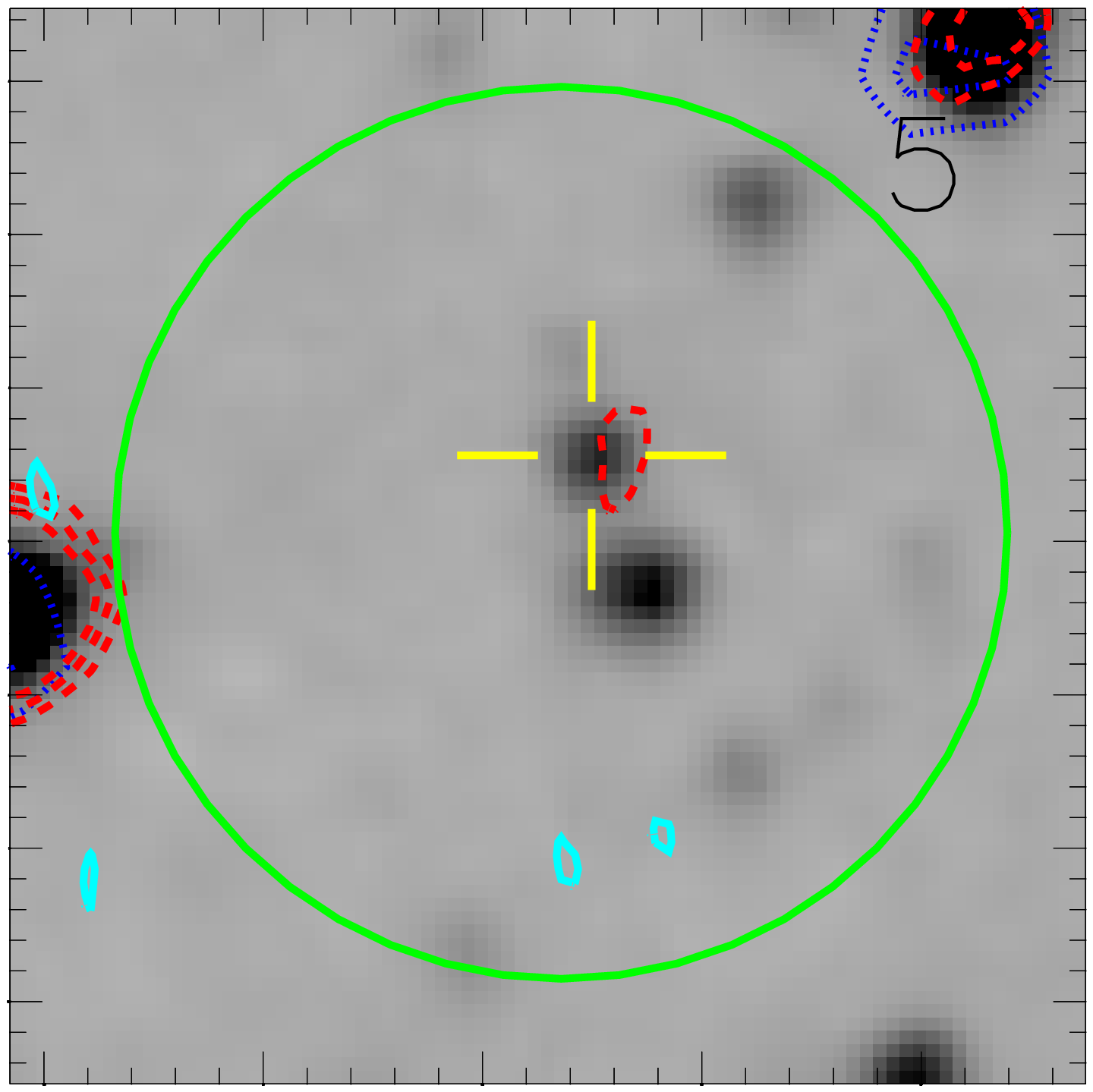}\includegraphics[width=0.15\textwidth]{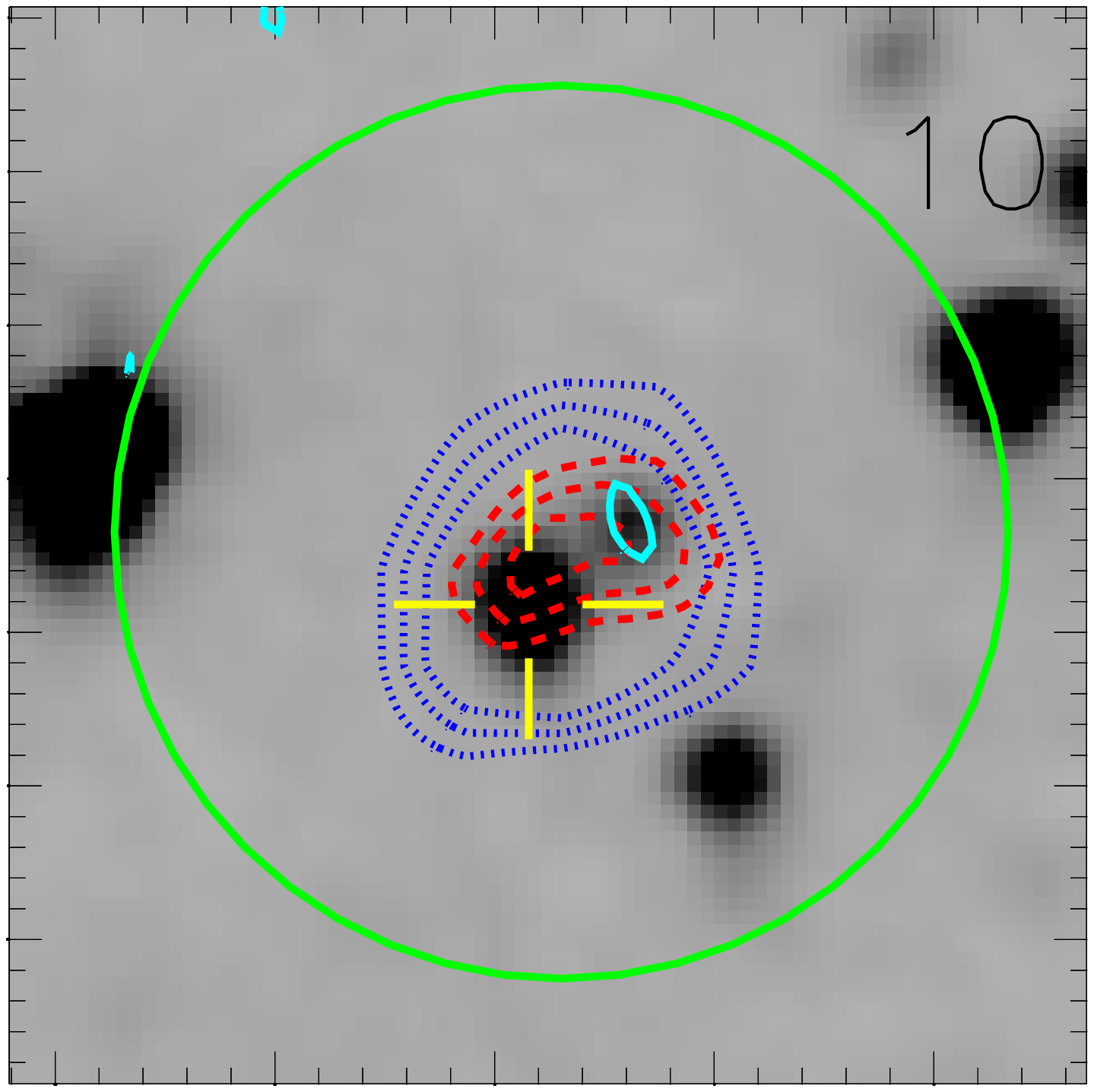} \hspace*{0.02\textwidth} \includegraphics[width=0.15\textwidth]{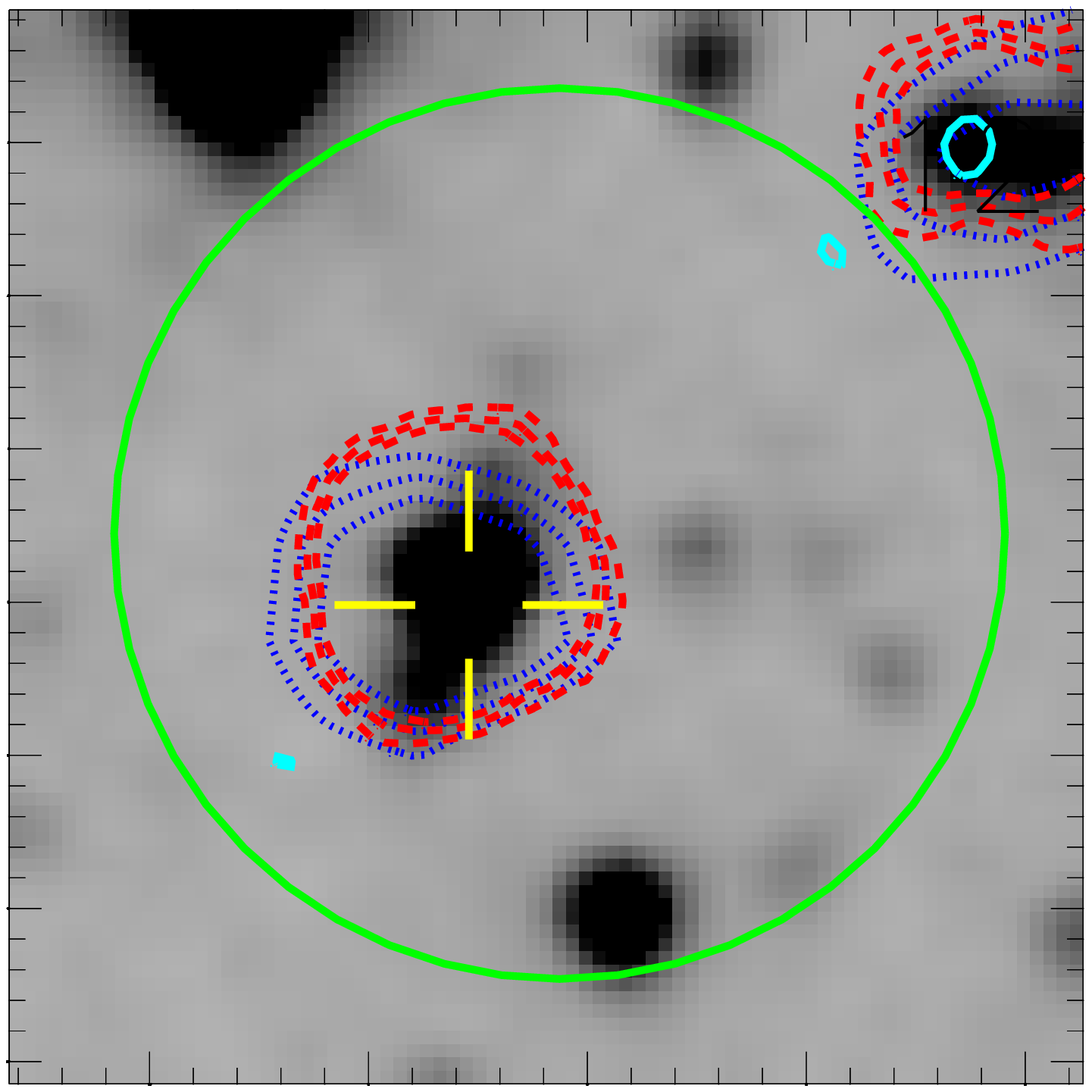}   \hspace*{0.15\textwidth} 

\caption{IRAC 3.6\,$\mu$m images of the eleven 850\,$\mu$m detected SMGs
  in our sample centered on the positions in Table~1. The images are separated into three groups according
  to their projected distance to the cluster center: $r_c<$\,0.25\,Mpc
  (left), $r_c=$\,0.25--0.7\,Mpc (middle), and $r_c=$\,0.75--1.0\,Mpc
  (right).  We show the 850\,$\mu$m beam as a green circle, and the
  identified counterparts with the yellow crosses.  We overlay [3, 4,
    5]\,$\sigma$ 24\,$\mu$m contours with a dashed red line, a
  4\,$\sigma$ contour at 1.4\,GHz in cyan, and [3, 4, 5]\,$\sigma$ for
  PACS 70\,$\mu$m in solid blue. The most likely counterparts are
  frequently radio detected and usually correspond to the brightest
  PACS counterpart in the beam.  The major tick marks show 5$\arcsec$
  increments. The white line at the bottom left of the first panel indicates 50\,kpc at $z=$\,1.46.   }
\label{fig:grids} 
\end{figure*}

\subsection{Radio Observations} \label{sec:data:radio}

We retrieve archival JVLA observations of XCS\,J2215 (Project
11A-241) taken in $L$-band in A-configuration over six nights in 2011
August to September.  For these observations the correlators were
configured to give two spectral windows centered at 1.264\,GHz and
1.392\,GHz, each containing 64 channels with 128\,MHz total
bandwidth.  The total on-source integration time was 15\,hrs. The
amplitude and passband calibrator 3C\,48 was observed at the beginning
of observation each night, and every two scans of XCS\,J2215 were
bracketed by the 2\,min scans of the phase calibrator J\,2246$-$1206.

\begin{figure*}
\includegraphics[width=0.15\textwidth]{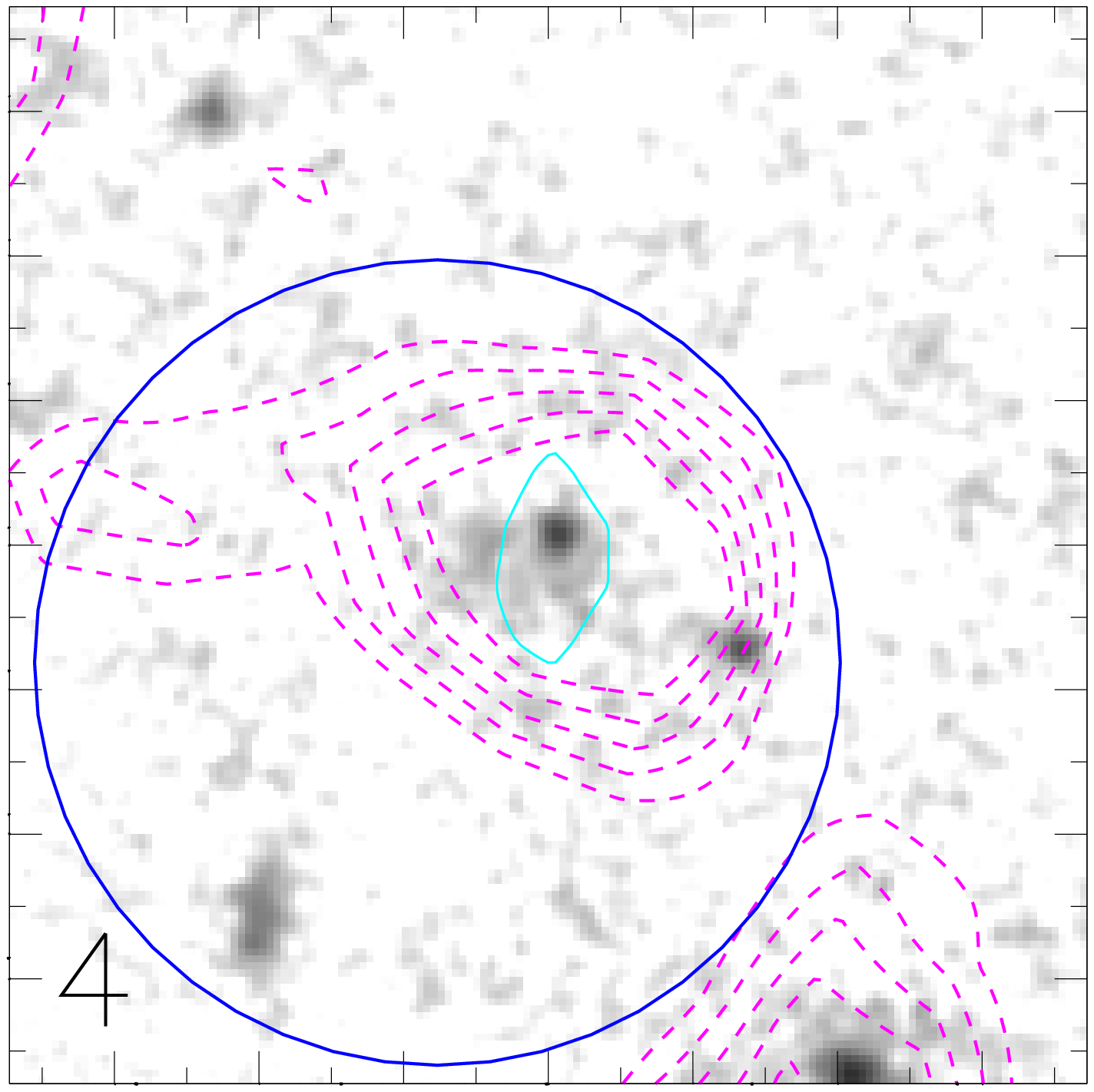} \includegraphics[width=0.15\textwidth]{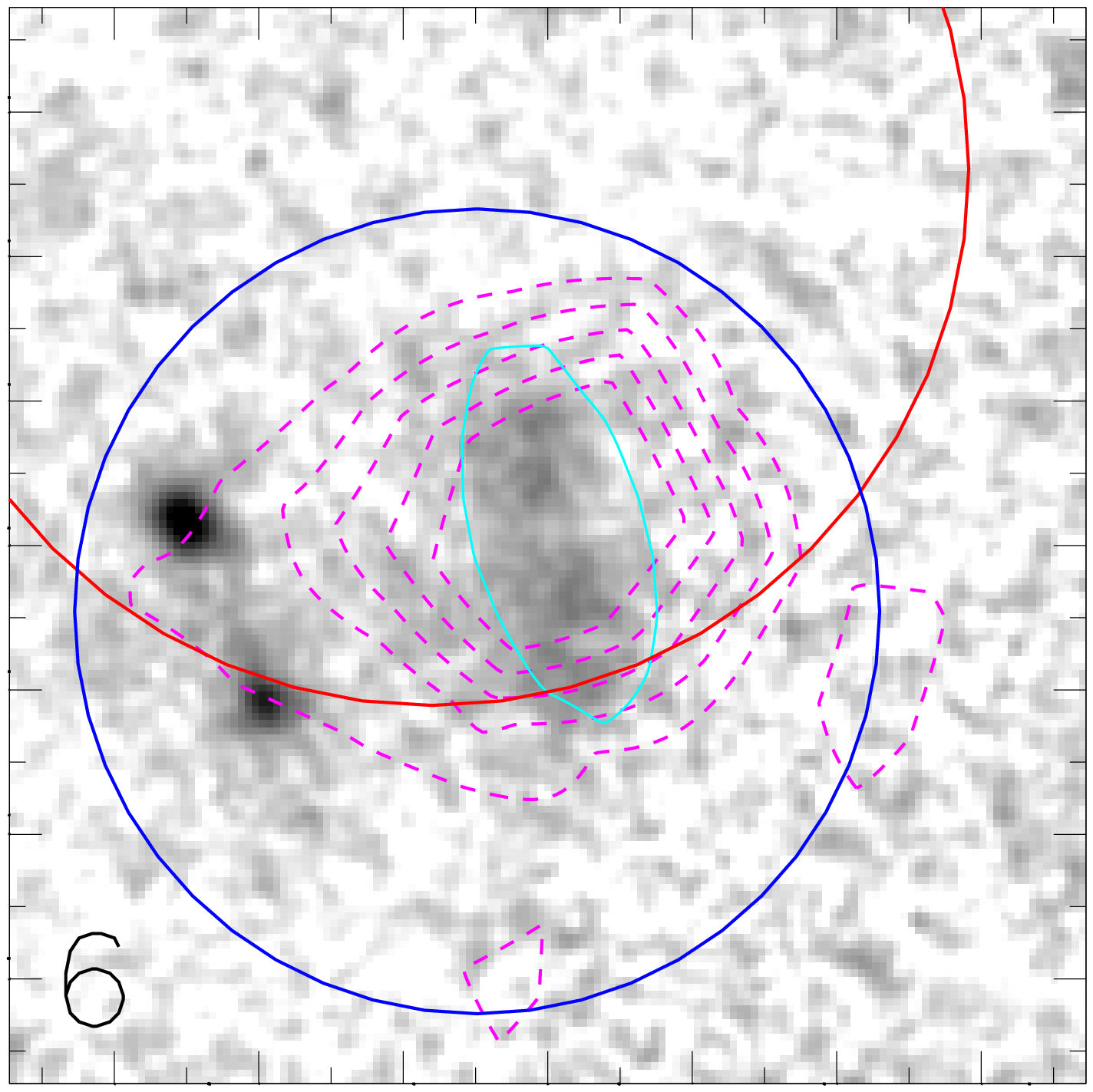}\hspace*{0.02\textwidth}
\includegraphics[width=0.15\textwidth]{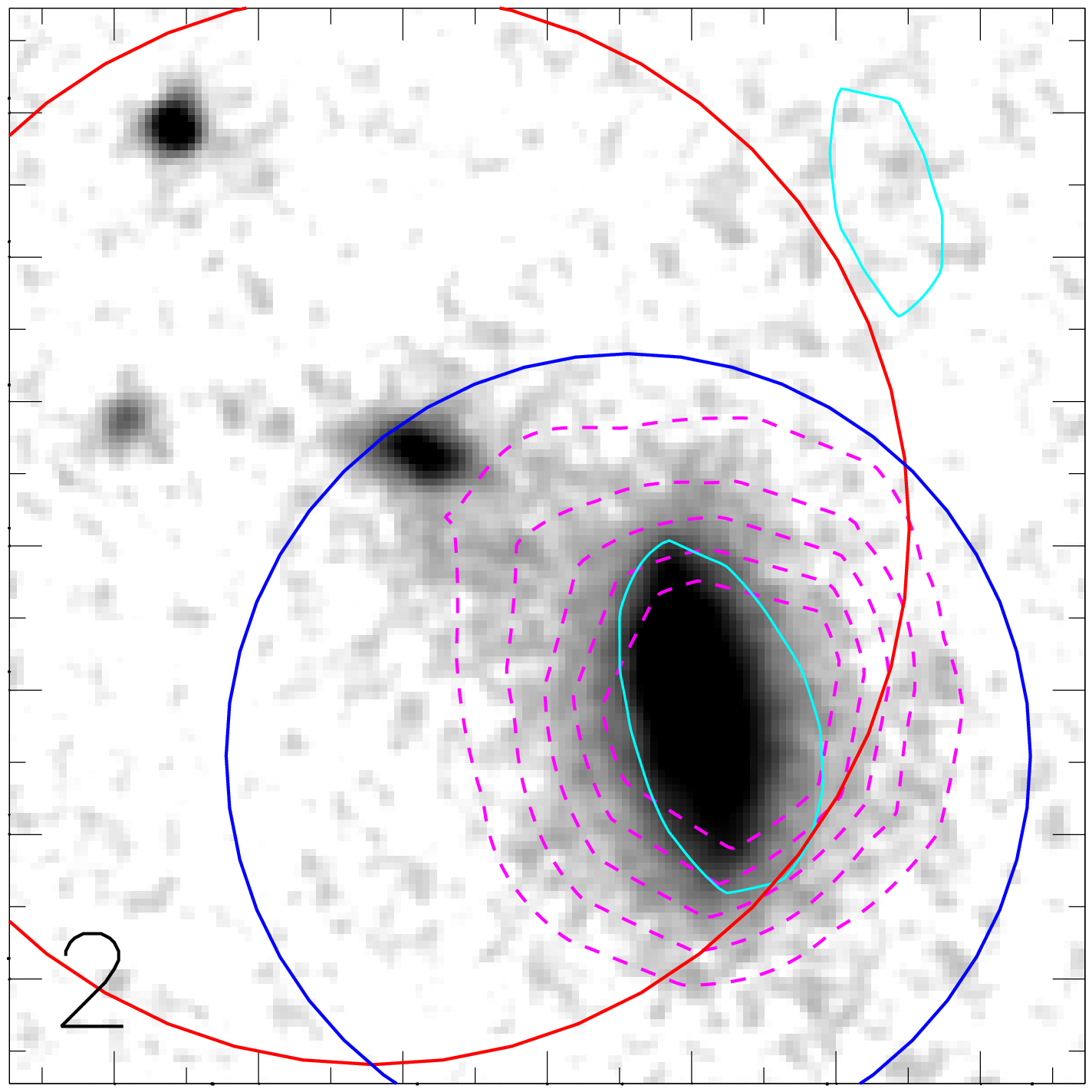}\includegraphics[width=0.15\textwidth]{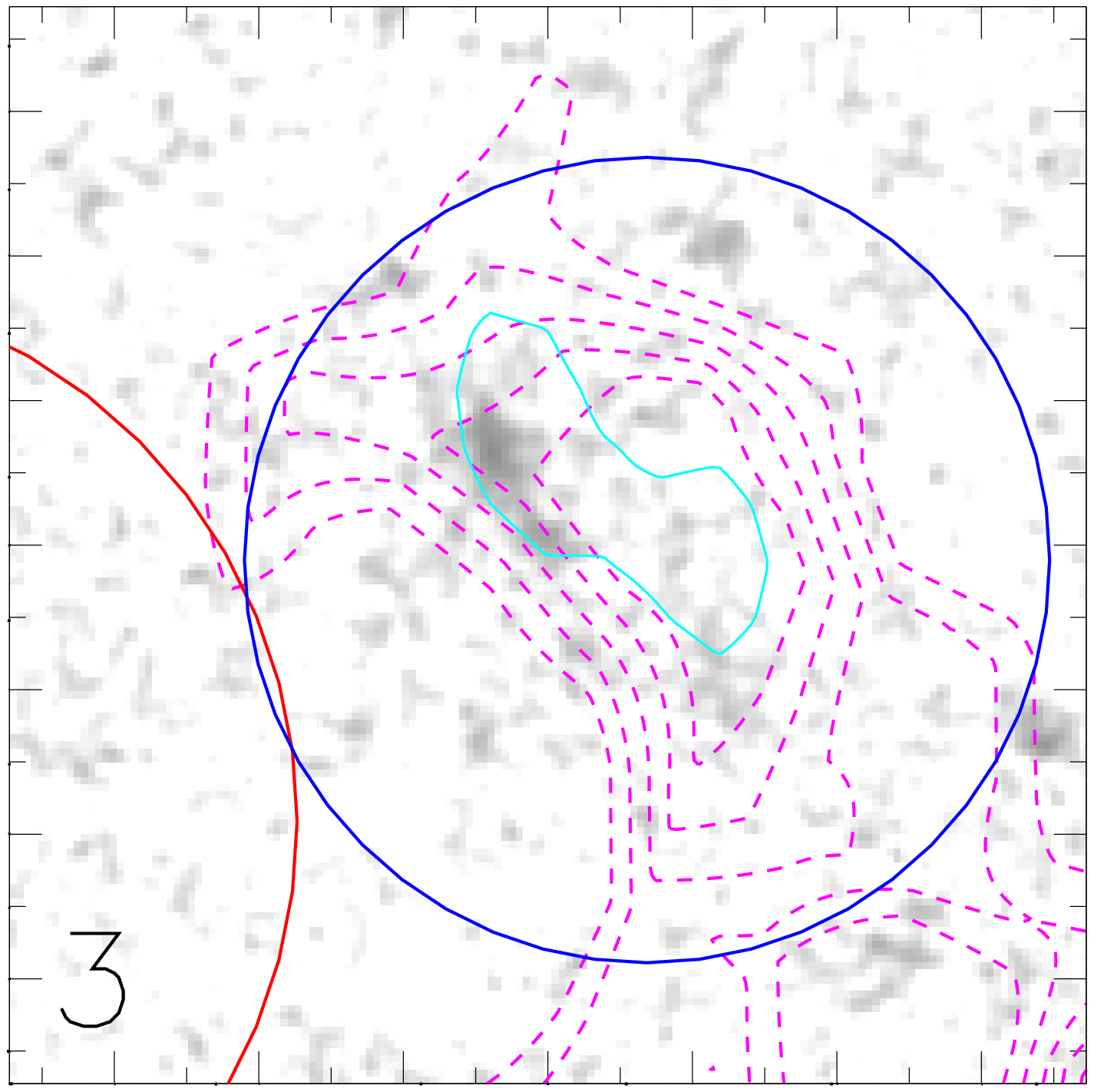} \hspace*{0.02\textwidth} \includegraphics[width=0.15\textwidth]{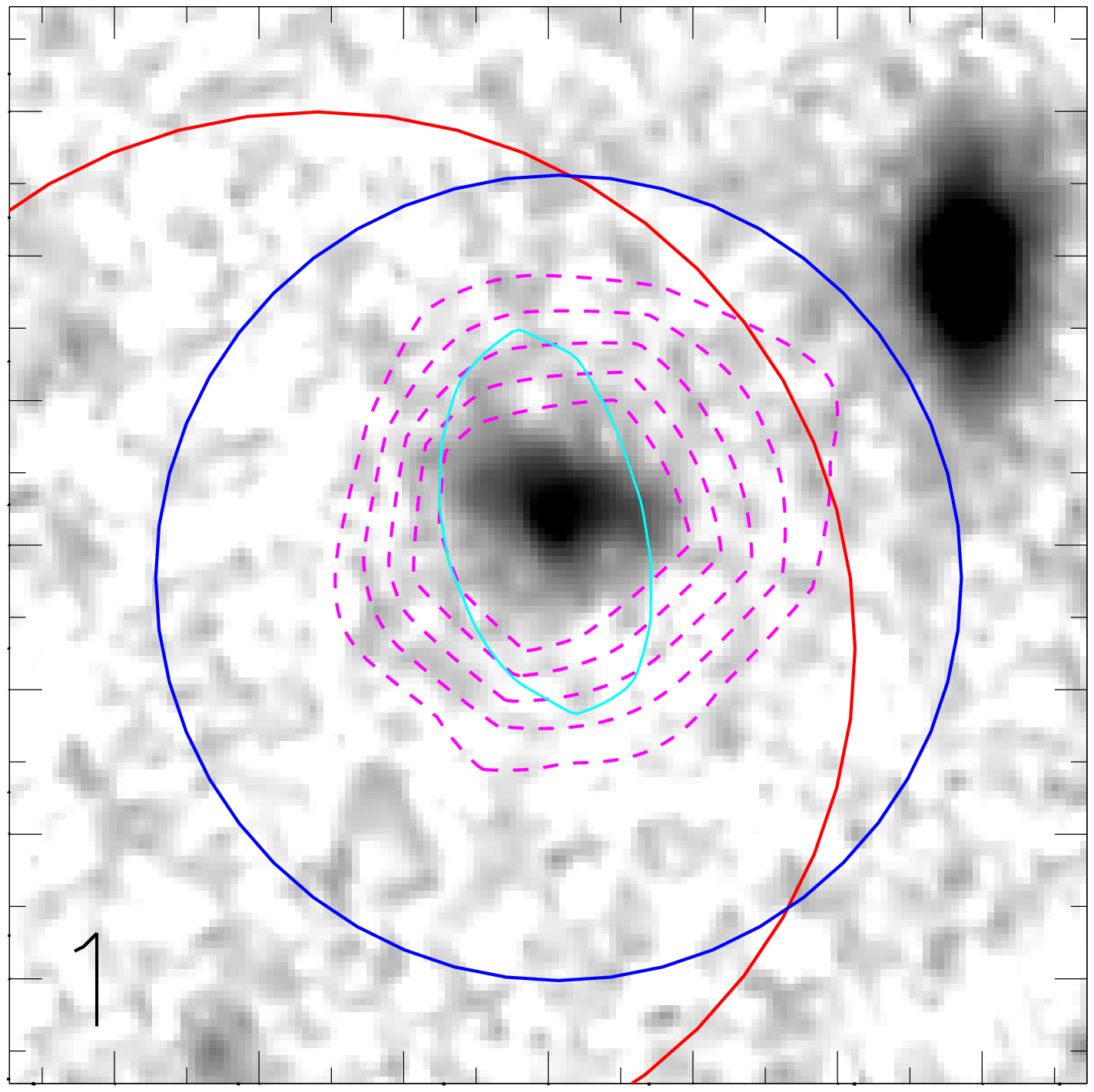} \includegraphics[width=0.15\textwidth]{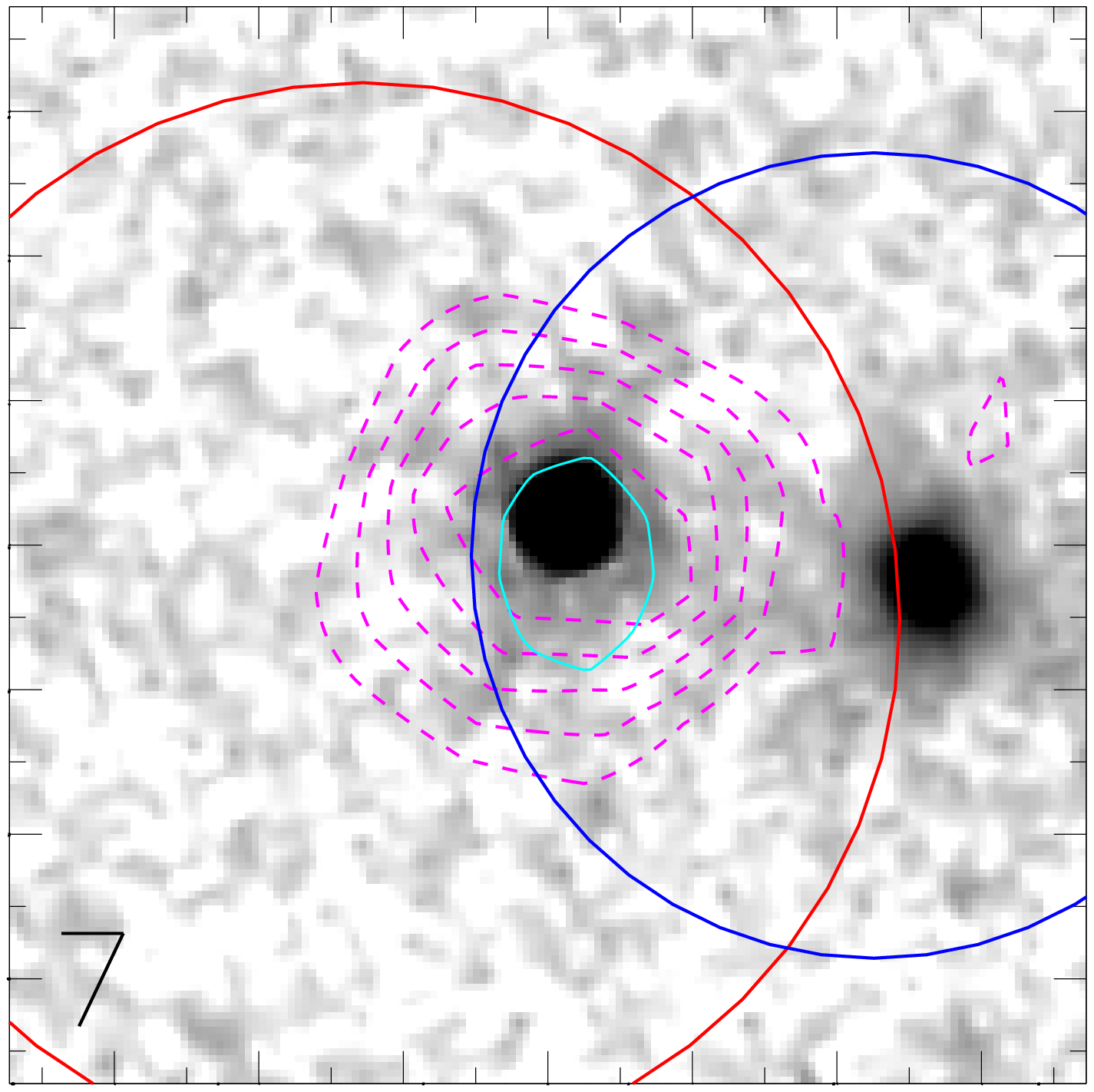}\\
\includegraphics[width=0.15\textwidth]{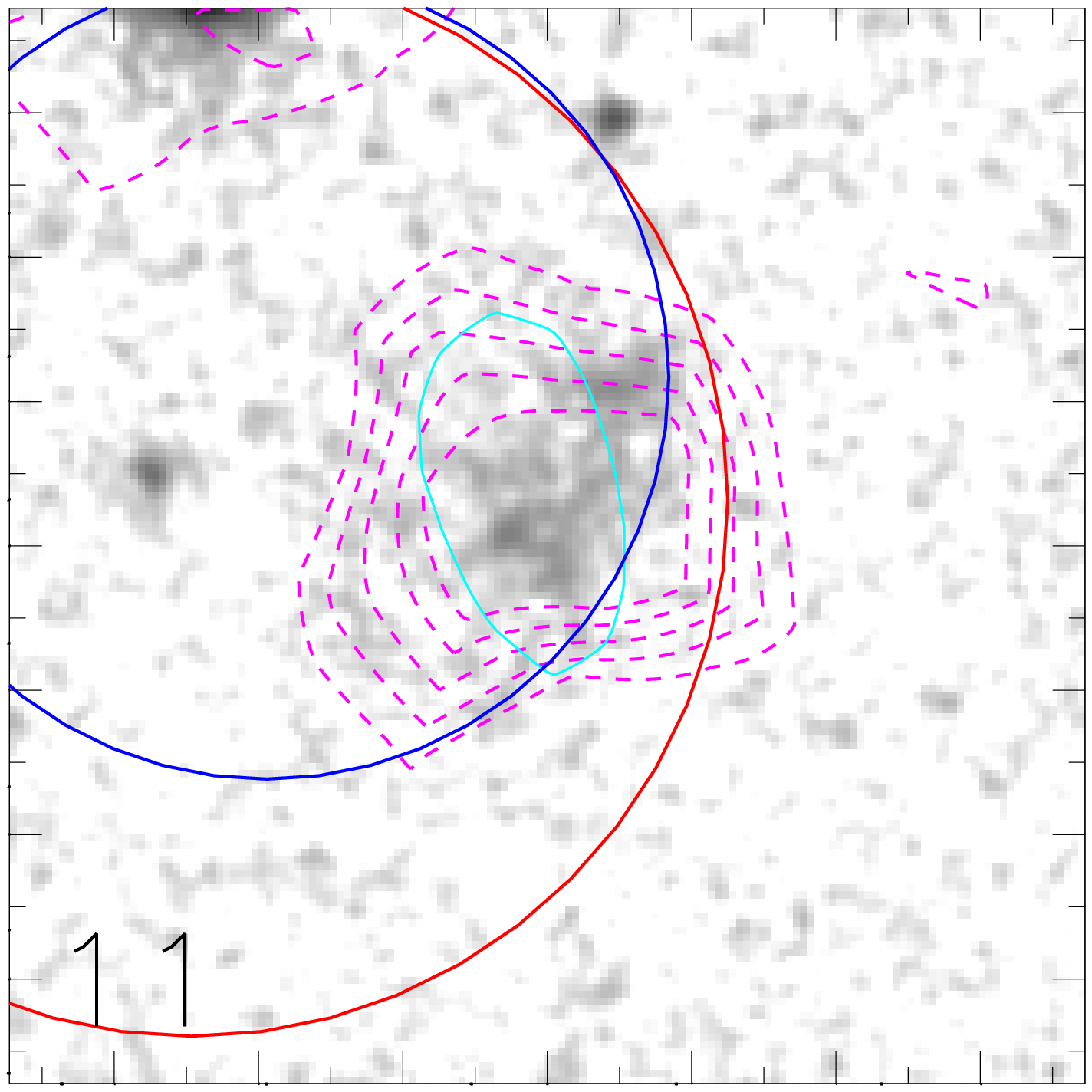} \includegraphics[width=0.15\textwidth]{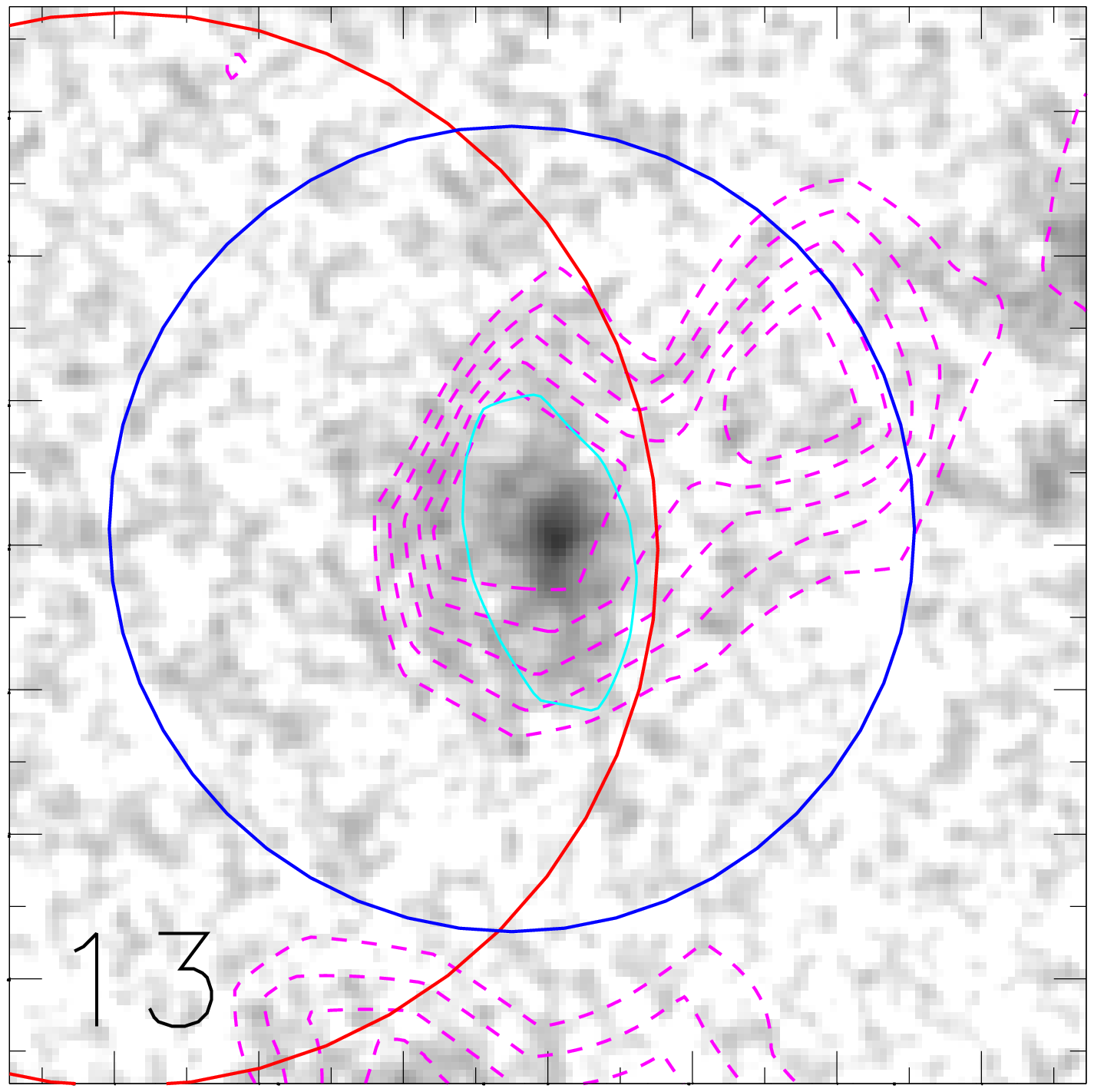}\hspace*{0.02\textwidth}
\includegraphics[width=0.15\textwidth]{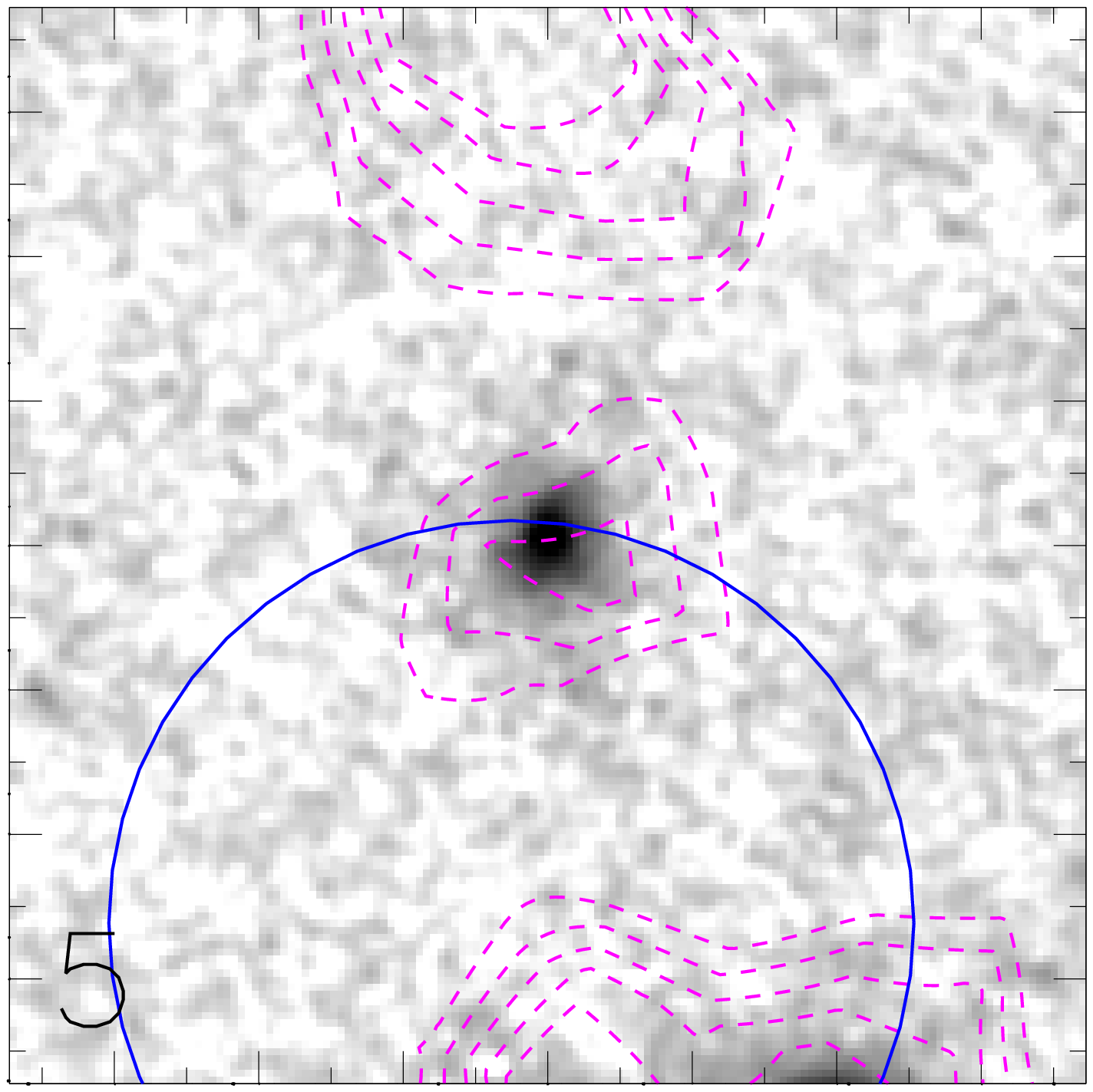}\includegraphics[width=0.15\textwidth]{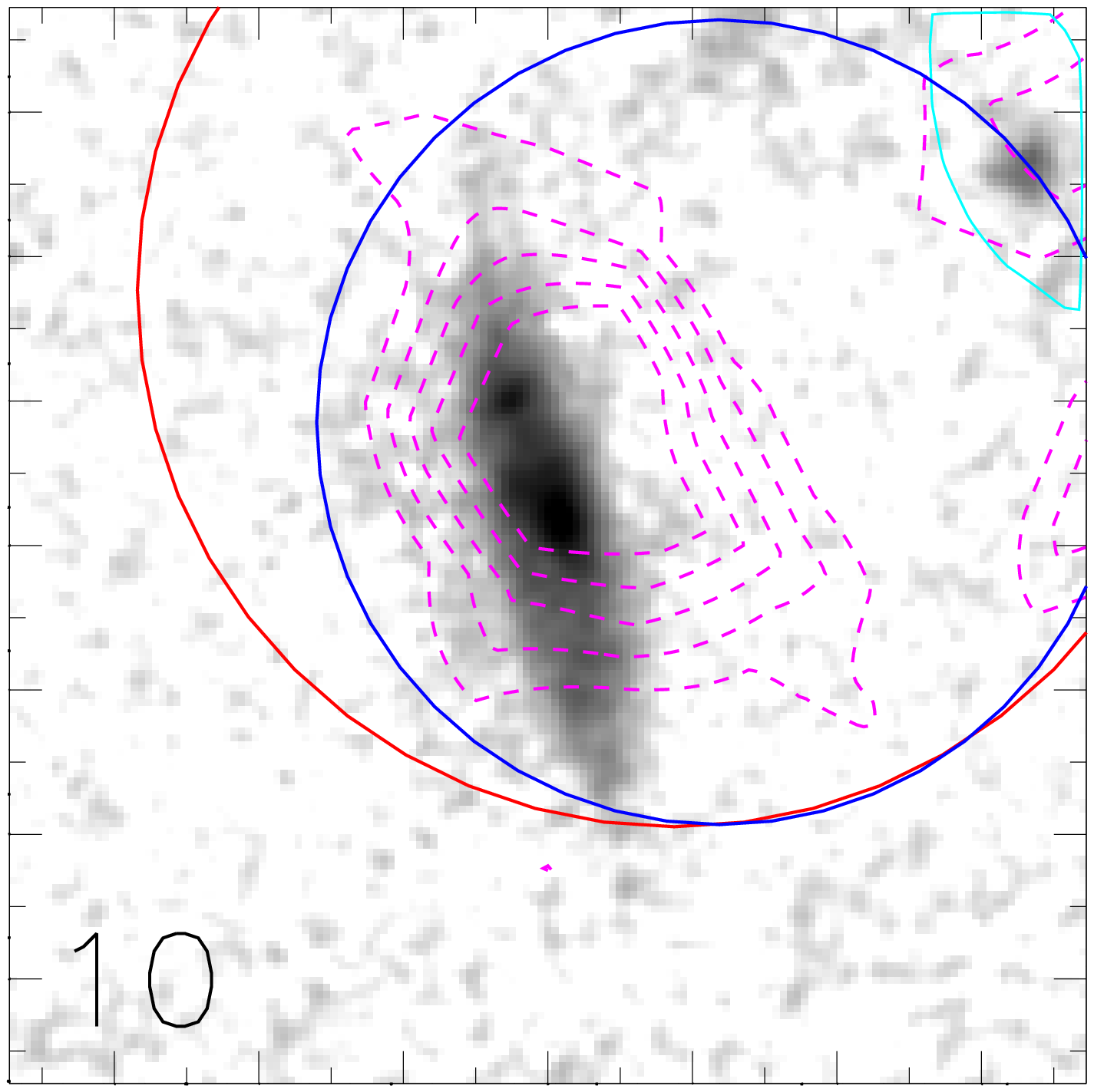} \hspace*{0.02\textwidth} \includegraphics[width=0.15\textwidth]{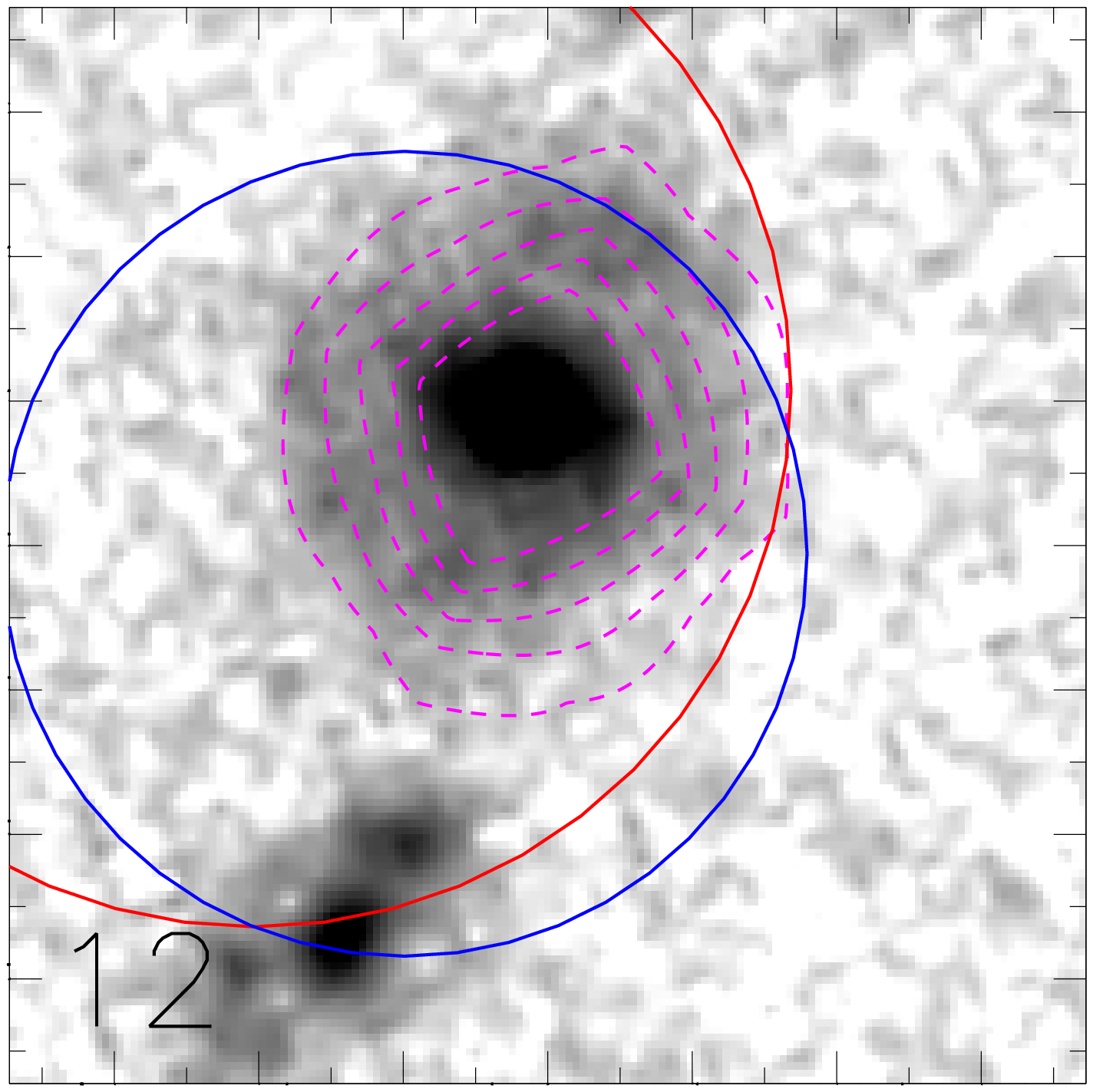}   \hspace*{0.15\textwidth} 

\caption{{\it HST} ACS F814W images of counterparts to eight of the 850\,$\mu$m
  detected SMGs in our sample and WFC3 F160W imaging of the remaining
  three (SMG\,01, 07 and 12), separated in terms of radius as in
  Fig.~\ref{fig:grids}.  We show the SCUBA-2 450\,$\mu$m beam as a red
  circle, the PACS 70\,$\mu$m as a blue circle, the IRAC 3.6\,$\mu$m as
  dashed magenta contours and the 4-$\sigma$ 1.4\,GHz radio contour in
  cyan.  The most likely counterparts from the radio/MIPS-based
  deblending are consistent with the positions of the 450\,$\mu$m
  sources (where present) and are also the brightest 8\,$\mu$m sources
  in the error circles, evidence of their very red infrared colors.
  The ACS images are $\log$-scaled and have been smoothed with a
  0.05$\arcsec$ gaussian.  The major tick marks indicate 1$\arcsec$
  increments.  }
\label{fig:grids_opt} 
\end{figure*}

We reduce the data using the {\sc Common Astronomy Software Applications}\footnote{http://casa.nrao.edu/}
({\sc casa}) software \citep{McMullin07}. One of the two spectral windows (1.264\,GHz)
suffered from severe radio frequency interference (RFI), and we apply
the automatic RFI excision mode ``rflag'' of {\sc flagdata} package in
{\sc casa}.  The data from each spectral window are then calibrated
in bandpass, phase, and amplitude using the calibrators respectively.
The combined data from the two spectral windows are used to make a
continuum image with robust weighting and a pixel size of
0.4$\arcsec$.  The image is self-calibrated to improve the
sensitivity, which reaches a 4-$\sigma$ limit of $\sim
$\,30\,$\mu$Jy\,beam$^{-1}$ at the center of the final image. The
synthesized beam size is 2.3$\arcsec \times$\,1.2$\arcsec$
(PA\,$=$\,195\,degrees). Fig.~\ref{fig:grids} and
\ref{fig:grids_opt} show the contours of the radio emission over the infrared
and optical imaging.

%
%fig 3
%

\subsection{Archival Observations}
\label{sec:data:IR}

To investigate the restframe multiwavelength properties of the 
sub-millimeter sources seen in the cluster we have exploited the images and
photometry from {\it Hubble Space Telescope} ({\it HST}), Subaru and {\it Spitzer} compiled by
\citet{hilton10} \citep[see also][]{hilton09,dawson09}.  We show the Subaru $K_s$ and {\it Spitzer} IRAC imaging of the cluster in
Fig.~\ref{fig:s850} and thumbnails of the {\it HST} imaging of the individual SMGs in
Figs.~\ref{fig:grids}. 

The details of $z_{850}$, $K_s$,  IRAC and MIPS data can be found in \citet{hilton09,hilton10} and the references therein, but we repeat the essential information here. \vspace{2mm}

\noindent {\it HST} $z_{850}$: The observation was performed using ACS  through the  F850LP filter \citep{dawson09}  (project ID: 10496). The total exposure time was 16.9\,ks, and the 5-$\sigma$ magnitude limit of the catalog in \citep{hilton09} is $z_{850}\sim26.0$.   We  retrieved the ACS images of the cluster from the {\it HST} archive\footnote{http://archive.stsci.edu/}, as well as a 0.7-ks WFC3 F160W near-infrared image (project ID: 12990).  These images provide $\sim$\,1\,kpc resolution imaging of the cluster members at restframe wavelengths approximately corresponding to the $U$ and $R$-bands.  These images are displayed in Fig.~\ref{fig:grids_opt} to illustrate the restframe UV--optical morphologies of the SMGs. \vspace{1.5mm}

\noindent Subaru $K_s$: The observation was performed using MOIRCS instrument \citep{hilton09}. The total integration time is 1.24\,ks, and the 5-$\sigma$ magnitude limit is $K_s\sim$\,24.3.  \vspace{1.5mm}

\noindent {\it Spitzer} IRAC and MIPS: A total of 1500\,s integration  was obtained with IRAC (Program 50333) and  1800\,s with MIPS.  The IRAC photometry were measured with 4$\arcsec$ diameter circular aperture with aperture correction measured by \citet{barmby08}.  We use the IRAC source catalogue compiled in \citet{hilton10}, of which sources are cross-matched against their $K_s$ band catalogue to find the nearest match within 1.2$\arcsec$ radius.  Our 24\,$\mu$m catalogue also comes from \citet{hilton10} and contains 152 sources detected at $>$\,5$\sigma$, of which eight are  associated with cluster members (four spectroscopically confirmed and four selected using photometric redshifts). The 50\% completeness limit of the catalogue is $\sim$\,70${\mu}$Jy \citep{hilton10}, corresponding to a limit of $L_{IR}>$\,3$\times10^{11}L_{\odot}$ at the cluster redshift using templates from \citet{dale02} or a SFR of $>$\,60\,M$_{\odot}$\,yr$^{-1}$ based on  \citet{kennicutt98}.     \vspace{2mm}

In addition to the data described above, we also obtain longer-wavelength archival observations of the cluster taken with the PACS instrument on {\it Herschel}  at 70\,$\mu$m and 160\,$\mu$m. The two observations are performed under an open time OT1 program \citep[see][]{santos13} on 2012 April 30 (Obsid: 1342245177) and 2012 December 19 (Obsid: 1342257511) and are retrieved from the {\it Herschel} Science Archive.  We process the data using the standard procedures  with {\sc hipe} software \citep[build: 11.0.2938][]{ott10} \citep[see][for more details]{lutz11}.  We, then, use the {\sc sourceExtractorDaophot} routines in the package of {\sc hipe} \citep{ott10} to detect  sources, and measure their photometry in 4.2$\arcsec$ and 8.5$\arcsec$ radii apertures at 70\,$\mu$m  and 160\,$\mu$m respectively. The sky background  is measured in an annulus between 20$\arcsec$--30$\arcsec$ and an automatic aperture correction from {\sc sourceExtractorDaophot} is applied.  The 1-$\sigma$  noise in these images are 0.4\,mJy and 1.4\,mJy at 70\,$\mu$m and 160\,$\mu$m respectively, measured using random aperture photometry in  regions free from sources.  In the central 2$\arcmin$ area of the maps, we detect 23 and 14 sources at  70\,$\mu$m and 160\,$\mu$m respectively above a 3\,$\sigma$ limit. 
Following the discussion of astrometry in \S\ref{sec:data:SMG}, we confirm that the mean offset of sources in the 70\,$\mu$m map is insignificant ($\Delta{\rm RA}=-$0.1$\arcsec\pm$\,1.0$\arcsec$ and $\Delta{\rm  Dec}=$\,0.5$\arcsec\pm$\,1.3$\arcsec$) using the counterparts of sources at 24\,$\mu$m.

\subsection{Cluster Galaxy Catalogue and Redshift Measurements}\label{sec:results:redshifts}

In our analysis we use the catalogue of cluster galaxies compiled in \citet{hilton09} and \citet{hayashi14} \citep[see also][]{hilton07,hilton10,hayashi10,hayashi11}. Again, we only summarize the most relevant results here. 

The sample in \citet{hilton09} includes 64 member galaxies: 24 of them are spectroscopic confirmed, and the remainder are selected by photometric redshifts. The photometric redshifts ($z_p$) are calculated with the $i_{775}$, $z_{850}$, $J$, and $K_s$ photometry. The uncertainty  in the photometric redshifts, ${\delta}z = (z_s-z_p)/(1+z_s)$, of the 36 galaxies with spectroscopic redshifts at $z_s>$\,1.0, is $\sigma_{{\delta}z} = $\,0.039. The cluster members are selected by, firstly, requiring that their photometric redshift lies in $z_p=$\,1.27--1.65 (i.e.\ $z_p$ falls within the 2\,$\sigma_{{\delta}z}$ of the cluster redshift, $z=$\,1.46), and, secondly, applying a quality cut on the photometric redshift such that $p_{{\Delta}z}>$\,0.9 \citep[see][]{hilton09}. Using these two criteria,
of the spectroscopically confirmed non-members, only two would be misclassified as members based on their photometric redshifts, i.e.\ $\sim$\,5\% of photometric redshift members may be non-members of the spectroscopic sample. In addition, four of the 24 spectroscopic confirmed cluster members with $z_s=$\,1.445--1.475 are missed by the photometric redshift selection ($\sim$\,80\% completeness. 

\citet{hayashi14} present a narrow-band survey to search for [O{\sc ii}] emitters in and around  XCS\,J2215 using  two filters (NB912 and NB921) with Suprime-Cam on Subaru. The sample includes 170 [O{\sc ii}] emitters at $z \sim $\,1.46 with a redshift  accuracy of  $\sigma((z-z_s)/(1+z_s)) = $\,0.002. The central wavelength and full-width half maximum of the NB912 (NB921) filters are 9139 (9196)\,\AA\ and 134 (132)\,\AA, equivalent to velocity difference of $\sim $\,2000\,km\,s$^{-1}$ or a redshift range of ${\delta}z = \pm$\,0.02 at $z=$\,1.46. To exclude the contamination of $H_{\alpha}$ or [O{\sc iii}] emissions at higher redshifts, they have applied a $BzK$ color selection illustrated in \citet{hayashi10}. The 5-$\sigma$ limiting magnitudes of the  narrow-band filter observations are 25.2 and 25.4, respectively, corresponding to SFR are 2.6 and 2.2\,M$_{\odot}$\,yr$^{-1}$  according to the relation of \citet{kennicutt98}. 

\section{Submm Sources Detections} \label{sec:results:id}

In the SCUBA-2 850\,$\mu$m map (Fig.~\ref{fig:s850}), we detect 16 sources above 4-$\sigma$ significance within the region with $>$\,50\% sensitivity, i.e.\ the radius of $\sim $\,5.4$\arcmin$ from the map center. The 4-$\sigma$ significance cut corresponds to a flux limit of $\sim $\,2.6\,mJy at the map center.  The reported fluxes are not statistically de-boosted, but we derive deblended fluxes for these below.  Eight of these 16 sources are located within 1\,Mpc of the cluster center (1.95$\arcmin$ radius), where the 1-$\sigma$ noise at 850\,$\mu$m is in the range of 0.63--0.80\,mJy\,beam$^{-1}$.  To estimate the false detection rate we search for negative ``sources'' in the map and confirm that there are no negative ``sources'' with S/N\,$<-$\,4 in this area.  

Using the number counts of the 870\,$\mu$m sources in the Extended {\it Chandra} Deep Field South submillimeter survey in \citet{karim13}, we  derive the density of sub-millimeter sources with a flux limit of $\sim 2.6$\,mJy, and estimate that the expected number in a typical field is $\sim $\,2 within the same projected area. Thus, the central regions of the XCS\,J2215 map display a $\sim$\,3--4\,$\times$ overdensity in terms of sub-millimeter sources.  

In addition, there are eight fainter 850\,$\mu$m sources with S/N\,$=$\,3--4  in the central 1\,Mpc area. Since many of these modest significance sources may be spurious, we use the data at 450, 160, and 70\,$\mu$m to verify the detections, and find that three of the eight are also simultaneously detected  with S/N\,$>$\,3 at all three of the shorter wavelengths.    This detection fraction is
consistent with a simple test of the false detection rate of faint 850\,$\mu$m sources from searching for negative ``sources'' in the  map,  which indicates a false detection rate of about
70\% for sources with S/N\,$=$\,3--4.  We also verify that none of these negative ``sources'' are detected at S/N\,$>$\,3 in all three of the shorter wavelength bands, which suggests that the three
faint 850\,$\mu$m sources with S/N\,$=$\,3--4 which are detected in MIPS/PACS are likely to be real.  Therefore, we include these three sources (SMG\,11, 12 and 13) into the final catalogue, and discard the other five. 

To summarize,  we list the fluxes from mid-infrared to radio of the eleven sources in Table~\ref{table:850srcs}, and show
the IRAC 3.6\,$\mu$m thumbnails of each in Fig.~\ref{fig:grids}. The ID of the sources are numbered in descending order of 850\,$\mu$m S/N.

\section{Counterpart Identifications}

To precisely locate the sub-millimeter sources we first use both the high-resolution 24\,$\mu$m and 1.4\,GHz images to identify their counterparts, and then match these to  sources in the optical and near-infrared data. Next, we use the redshift catalogue of Hilton et al.\ and the  [O{\sc ii}] emitter sample of Hayahi et al.\ to determine if the counterparts are likely cluster members. 

As shown in Fig.~\ref{fig:grids}, all eleven of the 850\,$\mu$m
sources have possible counterparts at 24\,$\mu$m.  Indeed, many of
them have multiple candidates within the 850\,$\mu$m SCUBA-2 beam
(Fig.~\ref{fig:grids}).  At 1.4\,GHz, we find unique counterparts above the 4-$\sigma$ radio flux limit ($\sim
$\,30\,$\mu$Jy) in the beam of six 850\,$\mu$m sources, and multiple counterparts in the three sources (SMG\,01, 02 and 03).

To determine the statistical significance of these
potential counterparts we calculate the standard corrected Poissonian
probability of a positional match \citep[$P$,][]{downes86}.  In
Table~\ref{table:850srcs}, we list the counterpart probabilities
calculated from the 1.4\,GHz and 24\,$\mu$m catalogues.  The
probabilities for all the proposed 1.4\,GHz identifications of
$P_{1.4}\leq $\,0.05 (normally taken to signify a robust
identification), and for those without 1.4\,GHz identifications, the
probabilities derived for the 24\,$\mu$m counterparts are also
$P_{24}\leq$\,0.05 (fortunately, the two 850\,$\mu$m sources which
lack 1.4\,GHz counterparts, SMG\,05 and SMG\,12, are associated with
single 24\,$\mu$m counterparts).  Thus, for all eleven of the SMGs we
identify single, robust counterparts at either 1.4\,GHz and/or
24\,$\mu$m to the 850\,$\mu$m sources.  The higher-resolution PACS
70\,$\mu$m and SCUBA-2 450\,$\mu$m imaging also provide an alternative
route to test these identifications and we show the PACS map in
Fig.~\ref{fig:grids} and the 70\,$\mu$m and 450\,$\mu$m beams in
Fig.~\ref{fig:grids_opt}, confirming the likely counterparts.

\subsection{Multiplicity and 850\,$\mu$m Flux Deblending}

Recent ALMA follow-up of sub-millimeter sources taken from single-dish
blank field surveys have shown that they frequently contain multiple
sub-millimeter sources when mapped at high spatial resolution
\citep{hodge13,simpson15}.  Indeed, \citet{hodge13} demonstrate that robust
counterparts (those with $P\leq$\,0.05) identified through radio and
mid-infrared emission, as we have done here, are correct in 80\% of
cases, but are also incomplete, recovering only 45\% of the
counterparts.  This suggests that our identifications are likely to
have reasonable purity, as indicated by their correspondence with the
70\,$\mu$m and 450\,$\mu$m positions.  As we see later, the fact that
many of these counterparts appear to be cluster members at
$z$\,=\,1.46 may also explain the high recovery rate of robust
counterparts.  This most likely reflects the lower-mean redshift of
these SMGs, $z\sim$\,1.5, compared to the bulk of the field SMG
population, $z\sim$\,2.5 \citep{simpson14}, which yields a more
advantageous $K$-correction in the MIPS and radio bands and hence a
higher recovery fraction.

%
%fig 4
%
\begin{figure*}
 \includegraphics[height=6cm,angle=0]{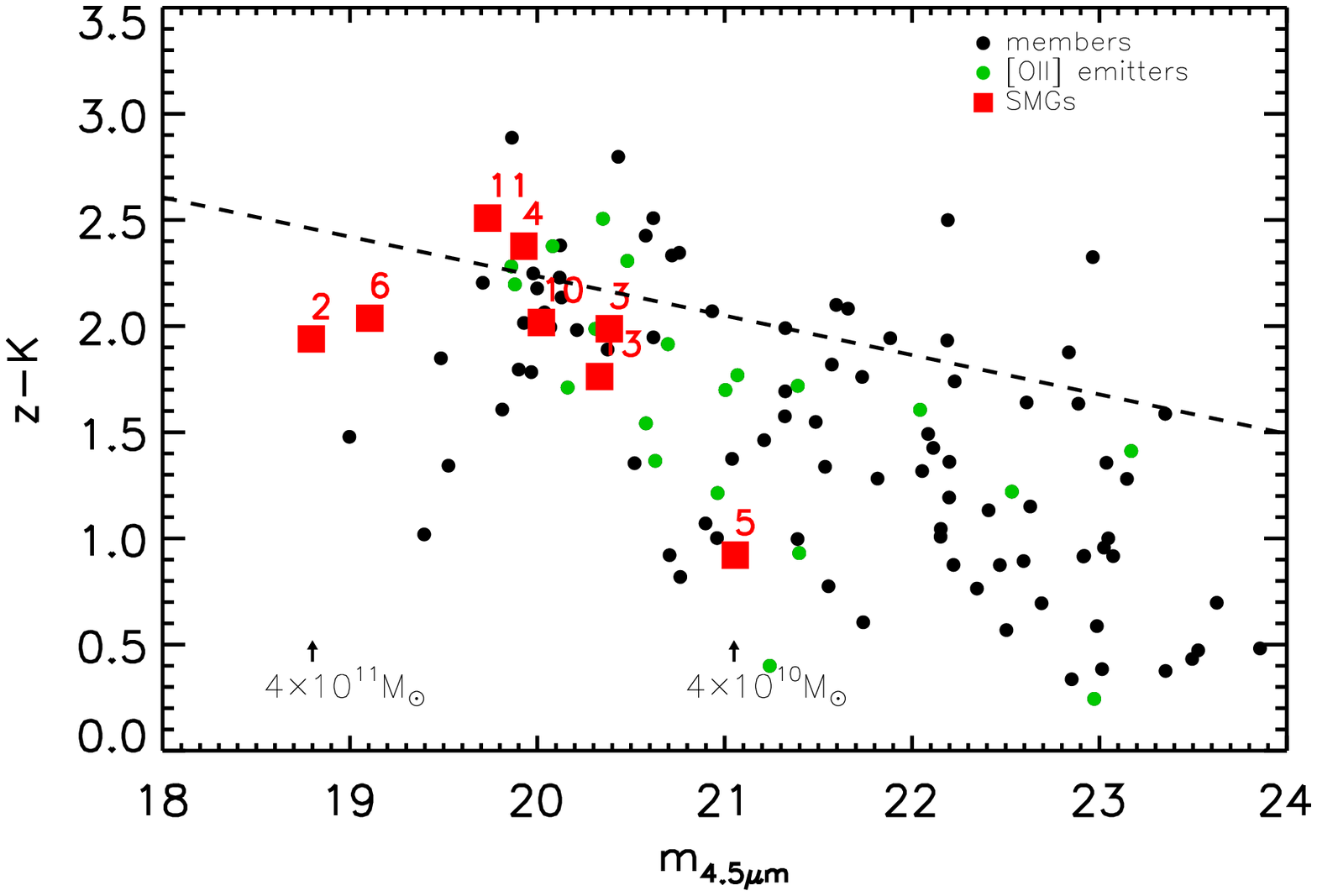} 
 \includegraphics[height=6cm]{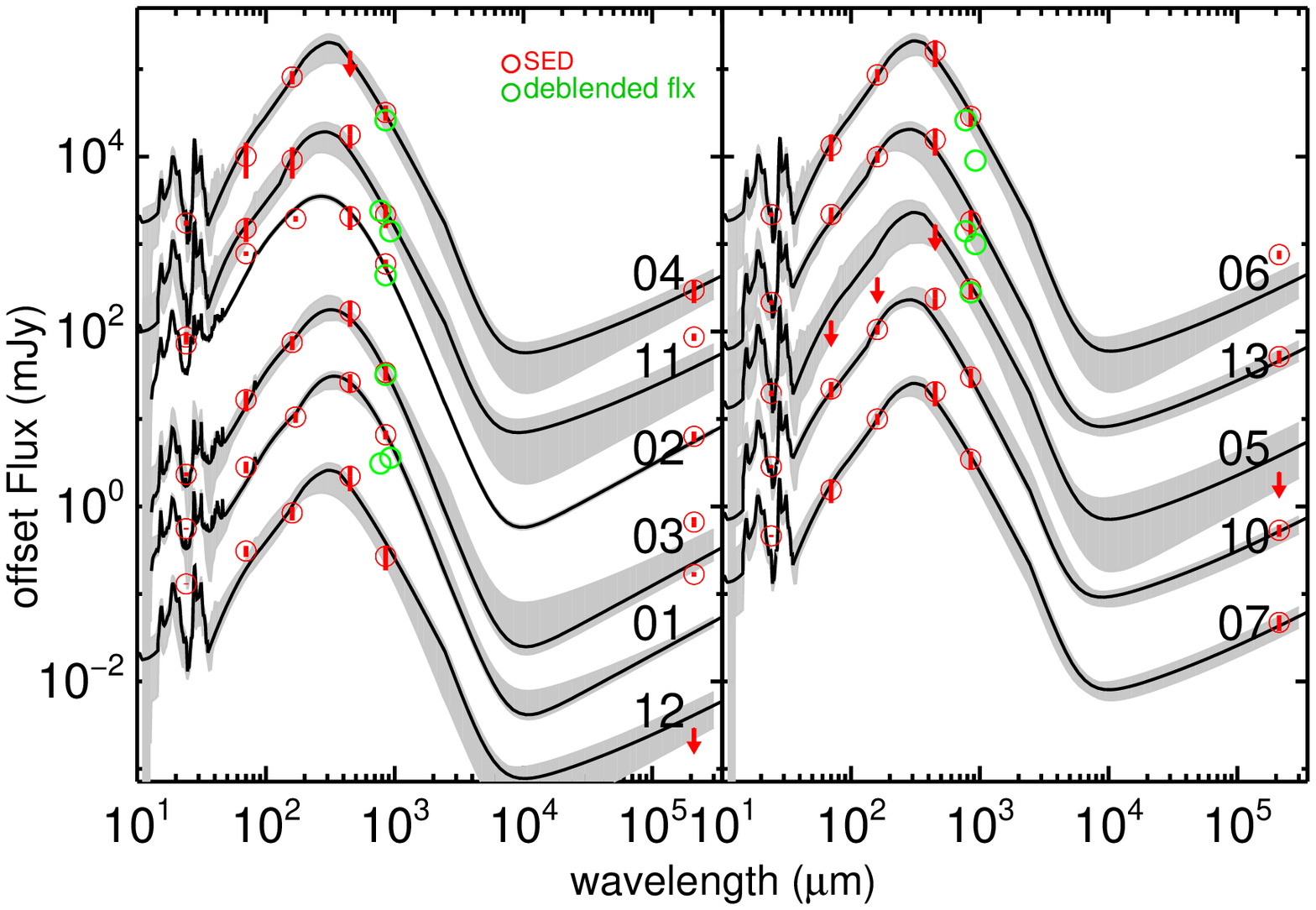} 
\caption{{\it Left:} The $(z'-K)$--m$_{\rm 4.5\mu m}$ color-magnitude
  diagram for cluster members (roughly corresponding to restframe
  $(U-I)$--M$_H$).  We plot the SMGs, the [O{\sc ii}]
  narrow-band emitters from \citet{hayashi10,hayashi14} and the
  photometric-redshift-members from \citet{hilton10}. The dashed line is the red sequence derived using the  early-type cluster members in \citet{hilton09}.
  This demonstrates that the SMGs are amongst the reddest cluster
  galaxies, likely due to  strong dust obscuration, and are also
  amongst the most luminous in the restframe $H$-band, suggesting they have high stellar mass. Similar behaviour has been seen in SMG
  members of other high-redshift clusters \citep[e.g.][]{smail14}. For reference, the stellar mass of the SMG\,02 and SMG\,05 calculated based on m$_{\rm 4.5\mu m}$ are noted in the bottom. {\it
    Right:} The infrared to radio SED of the SMG counterparts. The fluxes are plotted as red
  circles with 1-$\sigma$ errors and the non-detections are shown by
  3-$\sigma$ limits and we also show the
    deblended fluxes for the sources in green (see \S\ref{sec:results:sed}). The SED fit (solid line) excludes the 24\,$\mu$m,
  and the redshifts are fixed to the cluster redshift in all cases. The grey area shows the 1-$\sigma$ confidence
  range of the fit. 
  The majority of the SMGs have SEDs which are well-fit by the template
  libraries at the cluster redshift, showing that the dust emission from these galaxies is consistent with them being
  cluster members.  
}
\label{fig:sedfit} 
\end{figure*}

Fig.~\ref{fig:grids} also shows potential evidence of
multiplicity, with a number of SMGs appearing to have
multiple counterparts.  In the absence of sub-millimeter
interferometric fluxes for these sources, we have estimated
the sub-millimeter flux for each counterpart by deblending the
850\,$\mu$m map using the method in \citet{swinbank14} using a
position prior including all 1.4\,GHz and 24\,$\mu$m sources.  We
plot the fluxes of the individual counterparts (excluding any sources
with which are undetected based on the deblended fluxes), along with the spectral energy distributions (SEDs) of the
sources in Fig.~\ref{fig:sedfit}.  The fluxes of most sub-millimeter sources are
dominated by the counterpart selected by our probabilistic analysis,
and  their deblended fluxes are not significantly lower than the
integrated fluxes.  The only exceptions are SMG\,02 and SMG\,01.  For
SMG\,02, the deblended flux of the major counterpart is $\sim$\,2\,mJy
lower than the single dish flux of $S_{\rm
  850}$\,=\,5.9\,$\pm$\,0.7\,mJy.  For SMG\,01, the two possible
counterparts share the integrated 850\,$\mu$m flux evenly, and
correspond to two $\sim $\,3\,mJy SMGs.

\subsection{Cluster Membership}

Next, we match the radio and mid-infrared counterparts to the SMGs
with the galaxies detected in the IRAC 3.6\,$\mu$m image and the
photometric and spectroscopic redshift information which we have
associated with these galaxies.  Many of the matched galaxies are
cluster members based on spectroscopic redshifts
\citep[SMG\,06, 13:][]{hilton10}, [O{\sc ii}]\ emission at the
cluster redshift \citep[SMG\,03, 04, 11:][]{hayashi10,hayashi14}, or photometric redshifts
\citep[SMG\,02, 05:][]{hilton10}.  We tabulate the resulting
redshift information in Table~\ref{table:sfr}.  We note that all of the
SMG counterparts within 0.8\,Mpc radius of the cluster center appear
to be cluster members based on either  spectroscopic or photometric
redshifts \citep{hilton10}, or narrow-band [O{\sc ii}]\ emission
\citep{hayashi10,hayashi14}.  This suggests that gravitational lensing is
not a significant factor in the SMG over-density observed in the
cluster core.  However, we have no
redshift estimates for the most-likely counterparts of the three outer-most 850\,$\mu$m sources, SMG\,01, SMG\,07 or SMG\,12.

As we have noted, ALMA has demonstrated the need for robust
sub-millimeter identifications to reliably associate potential SMG
counterparts to sub-millimeter sources.  This is particularly true in
crowded fields such as the cluster core studied here, given the
density of potential counterparts, especially at 24\,$\mu$m.
Nevertheless, (in the absence of lensing) the strong over-density of sub-millimeter sources around 
XCS\,J2215 would suggest a
relatively low contamination rate (only two field SMGs are expected in
this area) and as we show in \S\ref{sec:results:sed} the SEDs of most of these counterparts are
consistent with those expected for ultra-luminous infrared galaxies
(ULIRGs) at the cluster redshift and so we suggest that the majority identifications are
likely to be correct.

\section{Analysis and Discussion} \label{sec:results}

%
%fig 5
%
\begin{figure*}
\includegraphics[height=6cm]{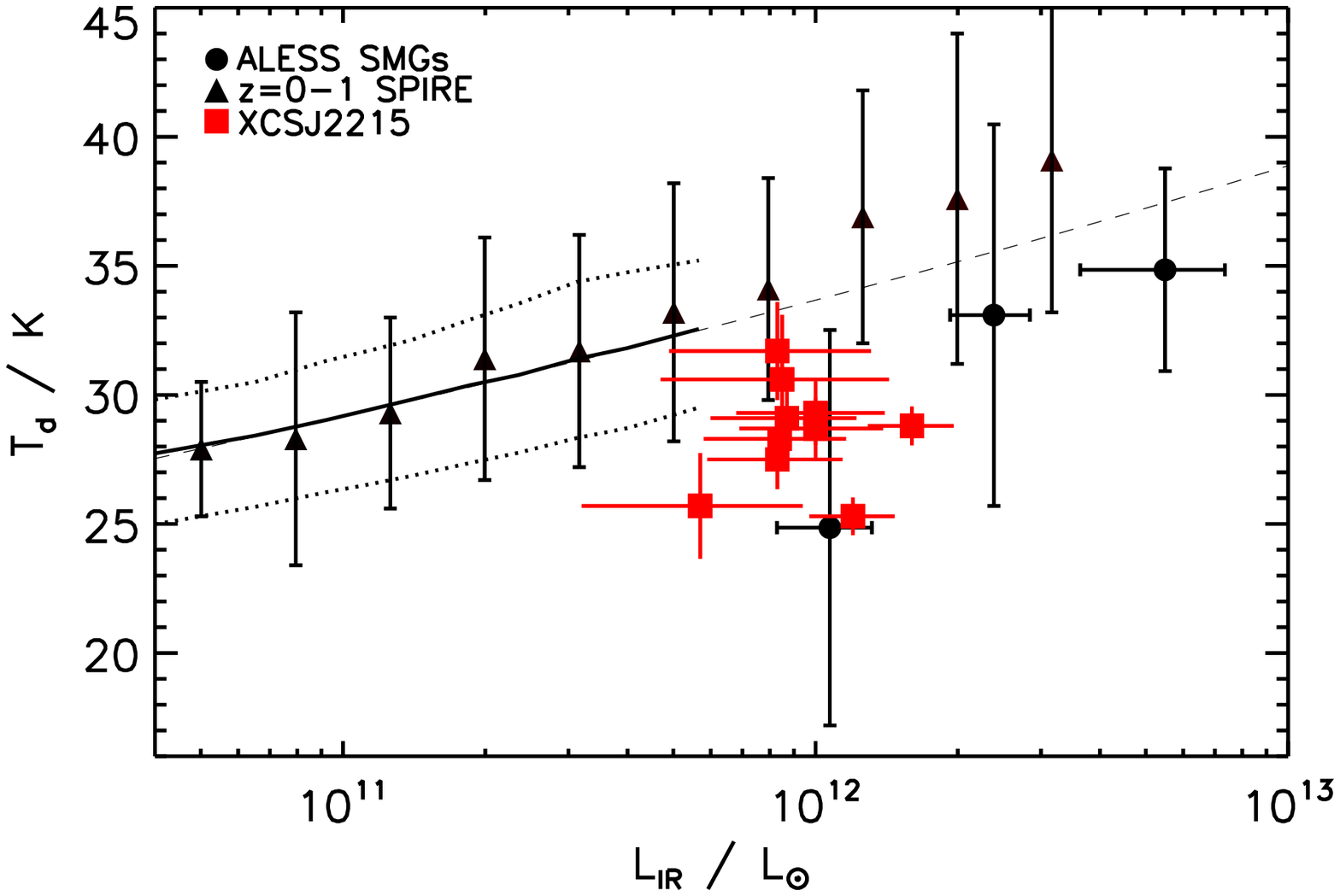} 
\includegraphics[height=6cm]{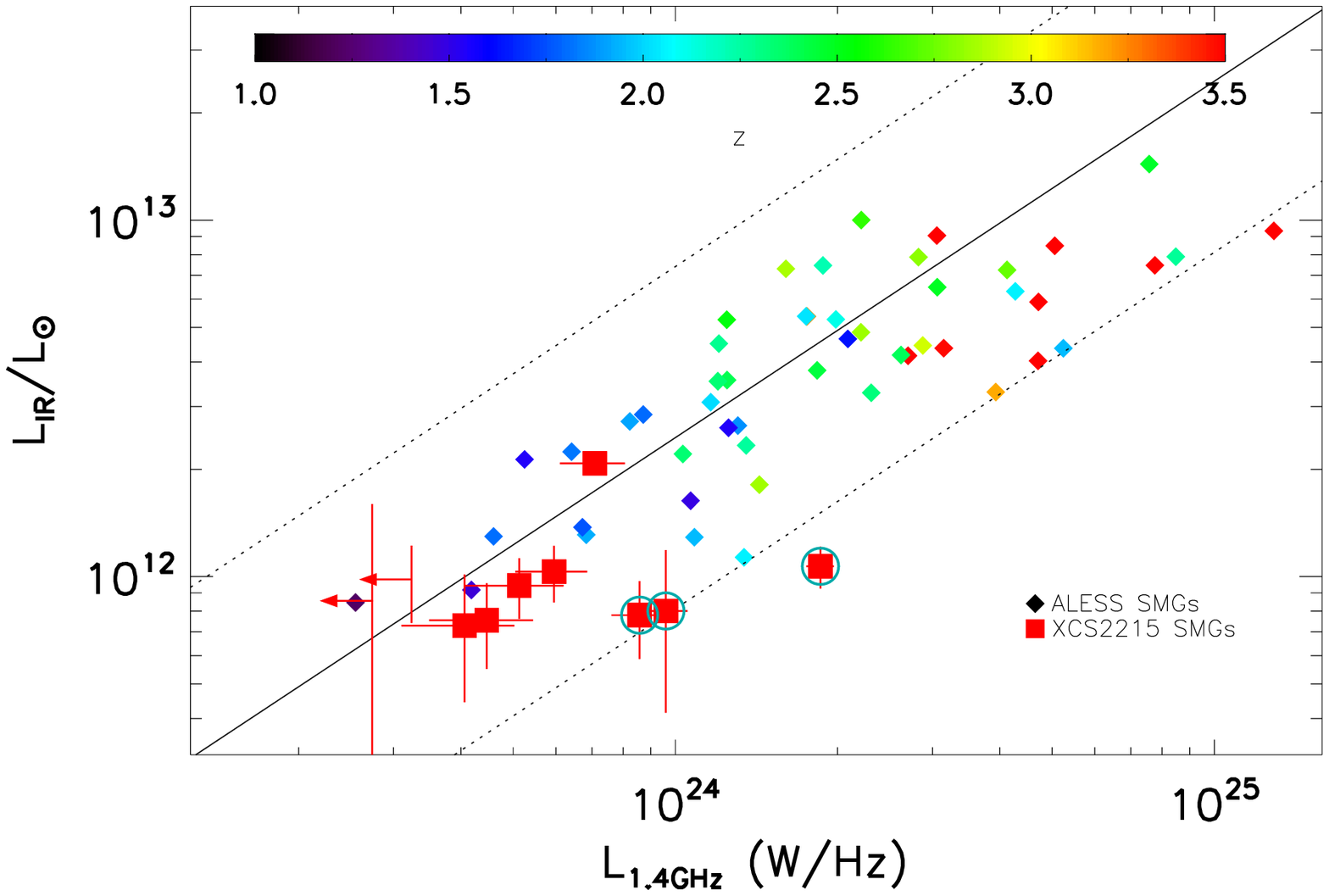}
\vspace{-4mm}
\caption{{\it Left:} Dust temperature and infrared luminosity of the
  SMGs within 1\,Mpc radius of XCS\,J2215 compared to low-redshift
  ULIRGs and high-redshift SMGs.  We include in the plot the $L_{\rm
    IR}$--$T_{\rm d}$ correlation derived for local 60\,$\mu$m selected
  {\it IRAS} galaxies by \citet{chapman03} \citep[solid line, with
    1-$\sigma$ dispersion shown by the dotted line; see
    also][]{chapin09}.  This plot shows that at a fixed $L_{\rm IR}$,
  the SMGs in XCS\,J2215 are $\sim $\,5\,K cooler than the average
  temperature of SPIRE-selected U/LIRGs at $z<$\,1
  \citep{symeonidis13}, and also $\sim $\,3--4\,K hotter than the
  median temperature of ALMA-identified SMGs with $z>$\,1.5 in the
  Extended {\it Chandra} Deep Field South \citep[ALESS,][]{swinbank14}.  This trend is consistent with a decline with redshift of the dust temperature at a fixed luminosity, suggestive of less compact starbursts at higher redshifts.  Note that the temperature are fitted with fixed redshifts and $\beta$, so that the uncertainty may be underestimated, see \S\ref{sec:results:sed} for the details. {\it
    Right:} The far-infrared--radio correlation for the XCS\,J2215
  SMGs comparing to the ALMA-identified SMGs
  from the ALESS survey \citep{thomson14}.  We only plot
  those ALESS SMGs detected at 1.4\,GHz in \citet{thomson14}, and they
  are color-coded according to their redshifts. The solid line shows the correlation from
  \citet{ivison10} with logarithmic infrared to radio flux ratio, $q_{\rm IR}=$\,2.40 and the dotted lines
  represent the 2-$\sigma_{q} = \pm0.48$ range. The three XCS\,J2215
  sources (SMG\,01, 06 and 11) with excess radio power in
  Fig.~\ref{fig:sedfit} are the three circled points. The ratio of far-infrared to radio luminosities  in the XCS\,J2215 sample is comparable to that seen in the radio-detected ALESS population at similar redshifts. }
\label{fig:lfir_td} 
\end{figure*}

\subsection{Spectral Energy Distributions} \label{sec:results:sed}

In Fig.~\ref{fig:sedfit} we show the $(z'-K)$--m$_{\rm 4.5\mu m}$
color-magnitude diagram for the cluster members, including the SMG
counterparts, [O{\sc ii}]-detected members from \citet{hayashi10,hayashi14} and
$K$-band members selected using photometric-redshifts from
\citet{hilton10}.  As seen in previous studies of SMG members of
high-redshift clusters \citep[e.g.][]{smail14}, and indeed the field SMG
population \citep[e.g.][]{simpson14}, the SMG counterparts tend to be both redder and more luminous than average field galaxies.  This is
consistent with their identification as massive, dusty starbursts,
where the dust extinction results in red UV--optical colors and the
combination of high mass and significant recent star formation means
that they are luminous sources in the restframe near-infrared. We
note that one counterpart, that for SMG\,05, appears fainter and bluer
than the rest, properties more consistent with the ``normal''
cluster population on the red-sequence.  This source is also unique in being undetected in all bands from 70\,$\mu$m to 1.4\,GHz, except at 850\,$\mu$m.

For the galaxies in Fig.~\ref{fig:sedfit} we can estimate the typical restframe $H$-band absolute magnitudes, which are
a crude tracer of their stellar masses,
by converting their observed 4.5\,$\mu$m (which samples
close to restframe $H$-band) using a model SED based on  the reddened burst
SED fit from \citet{simpson14}.  This predicts a combined $K$-correction and
distance modulus value of $-$44.9, which when applied to the observed 
4.5\,$\mu$m magnitudes gives a median restframe $H$-band absolute magnitude 
for the  SMG sample of $M_H=-$25.0\,$\pm$\,0.3.   This is marginally brighter
than the $M_H=-$24.6\,$\pm$\,0.2 derived for IRAC-detected ``field'' SMGs by \citet{simpson14} \citep[see also][]{hainline11},
even though the latter typically lie at higher redshifts.  The simplest interpretation
of this comparison is that the cluster SMGs we find in XCS\,J2215 may be marginally more massive than
comparably star-forming sources in the field.

We note that there are three sources which lie outside the coverage of the {\it HST} $z_{850}$ image and
so are missing from Fig.~\ref{fig:sedfit} and the calculation above.  However, all three appear to be AGN:
the counterparts to SMG\,07 and 12  in Fig.~\ref{fig:grids}  are both bright point-like components and
both of these counterparts are X-ray detected \citep{hilton10},  suggesting that these two SMGs host unobscured AGN
(which will contaminate their $H$-band luminosities). 
In addition, SMG\,01, appears to have a strong radio excess over that expected
for the best fit template SED in Fig.~\ref{fig:sedfit} (see \S\ref{sec:results:radio}), also suggesting an AGN component.  
Thus, intriguingly,  all three SMGs on the outskirts of the cluster host AGN.

\subsubsection{Template Fitting}

To derive the far-infrared luminosity and SFRs of the
850\,$\mu$m-detected cluster members, we  follow the procedure in \citet{swinbank14} and fit their SEDs using a
library of galaxy templates from \citet{chary01}, \citet{rieke09}, and
\citet{draine07}.  The fitting is performed with the photometry
from far-infrared to 1.4\,GHz, but excluding the 24\,$\mu$m band to
avoid the complexity caused by the redshifted PAH and Silicate
features at the cluster redshift.  The redshifts used in the fitting
procedure are fixed at the cluster redshift for all sources.
The results of this are shown in Fig.~\ref{fig:sedfit} and
Table~\ref{table:sfr}.
As Fig.~\ref{fig:sedfit} shows, the template libraries provide adequate
fits to the majority of the SMGs, indicating that their SEDs can be
well-described by standard templates at the cluster redshift.   The median of $\chi^2$ per degree of fredom (DOF) is 0.66, and the values are less than one for all sources except for SMG01, $\chi^2/DOF=$\,1.8; SMG02, $\chi^2/DOF=$\,1.6; and SMG12, $\chi^2/DOF=$\,2.8. 
If we allow the redshift for the sources without spectroscopic redshift to be a free parameter in the fits, the best fit redshifts are still located within the range of $z=$\,1.3--1.7, close to the cluster redshift. For all sources without precise redshifts the  fits with unconstrained redshifts are  not significantly better, $\delta\chi^2<$\,0.2, than those fixed at the cluster redshift. 

We note that the fitting of SMG\,05, which was already highlighted above, is much more weakly constrained  as it is based on a single detection at 850\,$\mu$m and limits. The best-fit template of SMG\,05 predicts fluxes above the
detection limits at 70\,$\mu$m, 160\,$\mu$m, 450\,$\mu$m and
1.4\,GHz.  This may imply that this source is in fact a
higher-redshift field SMG which is not related to the identified
cluster counterpart.  The presence of a field source within the sample would be
consistent from the predicted field contamination, but we note that our subsequent analysis of the
integrated SFR in the cluster does not change
significantly if this source is included or excluded.  Similarly, the
other three galaxies without precise redshifts (SMG\,01, 07 and 12),
are all located outside of $R_{200}\sim $\,0.8\,Mpc, and
so are not included in our estimate of the integrated SFR within the virial radius of cluster derived below. As a result  the
uncertainties over both their membership and the AGN contribution to their
far-infrared luminosities (see \S~\ref{sec:results:sed})  does not effect our
discussion.

\subsubsection{Infrared Luminosity and Dust Temperature}

Using the best-fit models for each SED, we calculate the infrared
luminosity $L_{\rm IR}$ in the wavelength range from 8--1000\,$\mu$m,
and the SFR from the relation in \citet{kennicutt98}
assuming a Salpeter initial mass function.  The results are listed in
Table~\ref{table:sfr}.  To compare the dust temperature in
these cluster galaxies with other SMGs, we also estimate the dust
temperature ($T_{\rm d}$) by fitting the fluxes from 160--850\,$\mu$m
using a grey-body model assuming a dust emissivity of $\beta=$\,1.5 and the redshifts in
Table~\ref{table:sfr}. We note  that it is likely that the uncertainties on the derived  temperature are underestimated as we have adopted 
both fixed redshifts and $\beta$ in the fitting.

 We show the distribution of $L_{\rm IR}$ and $T_{\rm d}$ for the
cluster SMGs in the left panel of Fig.~\ref{fig:lfir_td}, where we compare it to both
the $L_{\rm IR}$--$T_{\rm d}$ trends seen in samples of low-redshift ($z\sim0.4$) ULIRGs
from \citet{symeonidis13} and the median values for high-redshift
field SMGs from \citet{swinbank14}.  At the typical luminosities of
the SMGs in XCS\,2215, $L_{\rm IR}\sim $\,1\,$\times
10^{12}$\,L$_\odot$, the SMGs appear to have
temperatures which are $\sim$\,3--4\,K cooler than comparable luminous
low-redshift ULIRGs ($\sim $\,29\,K versus $\sim$\,34\,K respectively).  At
these luminosities the ALMA-identified field SMGs from
\citep{swinbank14}, which typically lie at higher redshifts
$z\sim$\,2--3, are even cooler ($\sim $\,25\,K).  This suggests that
the characteristic dust temperature of ULIRGs may decline with increasing
redshift, perhaps indicating a more extended distribution of dust in
SMGs, compared to local ULIRGs, although the influence of the
850\,$\mu$m sample selection complicates the interpretation of this
trend.  A number of other studies have suggested similar trends of
more extended dust and star-formation distributions in SMGs, using a
variety of observations \citep[e.g.][]{chapman03, hainline09, 
  menendez09, ivison10, swinbank14}.

\subsubsection{Different Measurements of Star Formation Rates}\label{sec:results:sfr}

 We compare the different tracers of star formation in the
  XCS\,J2215 galaxies in Table~\ref{table:sfr}.  Here we list the SFR
  calculated from the [O{\sc ii}] narrow-band fluxes
  \citep{hayashi10,hayashi14} and 24$\mu$m \citep{hilton10} for
  comparison to the far-infrared.  The SFR$_{\rm [OII]}$ of those SMGs
  detected as [O{\sc ii}] emitters are calculated according to the
  relation of \citet{kennicutt98}, which significantly underestimates
  the SFR of these dust-rich SMGs.  An extra attenuation factor of
  $\sim 3$ magnitudes at 3727\AA\ is required to bring the SFR$_{\rm [OII]}$ up to
  their SFR$_{\rm FIR}$. 
In addition, the non-detection of [O{\sc ii}] emission from the other
SMGs suggest an attenuation factor  of $> 4$ magnitudes.  To
compare the 24$\mu$m and far-infrared, we use
the tabulated SFR$_{\rm 24{\mu}m}$ fro SMG\,06 and 13 derived from the \citet{chary01} spectral
templates by \citet{hilton10} to calibrate the SFR of the other SMGs
from their 24\,$\mu$m fluxes.  As expected for the dusty SMGs we
find that  SFR$_{\rm 24{\mu}m}$ are not as biased as the SFR$_{\rm
  [OII]}$, although there are obvious biases associated with
mid-infrared-bright AGN in the sample and overall the large variation relative to the SFR$_{\rm FIR}$ may be explained by the contamination of PAH features.

\subsubsection{Radio Power}\label{sec:results:radio}

The right panel of Fig.~\ref{fig:lfir_td} compares the ratio of far-infrared and radio luminosities for the candidate $z=$\,1.46 cluster SMGs to those of a sample of radio-detected, ALMA-identified field SMGs drawn from \citet{thomson14}. 
We restrict ourselves to the radio-detected field sample to  mimic the requirement for radio identifications of counterparts in XCS\,J2215.
For the XCS\,J2215 SMGs the radio power
  density at rest-frame 1.4\,GHz is calculated assuming a spectral
  index of $\alpha=-$0.8.  We also plot   the far-infrared--radio
correlation from
  \citet{ivison10} and its 2-$\sigma$ boundaries and highlight the three XCS\,J2215
  SMGs (SMG\,01, 06 and 11) which exhibit excess radio power in
  Fig~\ref{fig:sedfit}.   Overall the XCS\,J2215 SMGs appear to have 
enhanced radio luminosities relative to their far-infrared emission, compared to the field population, although if we exclude the three radio-loud SMGs, the majority of the SMGs in XCS\,J2215 lie within the scatter of the far-infrared/radio luminosity ratios  of the  field SMG population.   

\subsection{Integrated star-formation rate}
\label{sec:results:SFR}

\begin{figure}
 \includegraphics[width=0.5\textwidth]{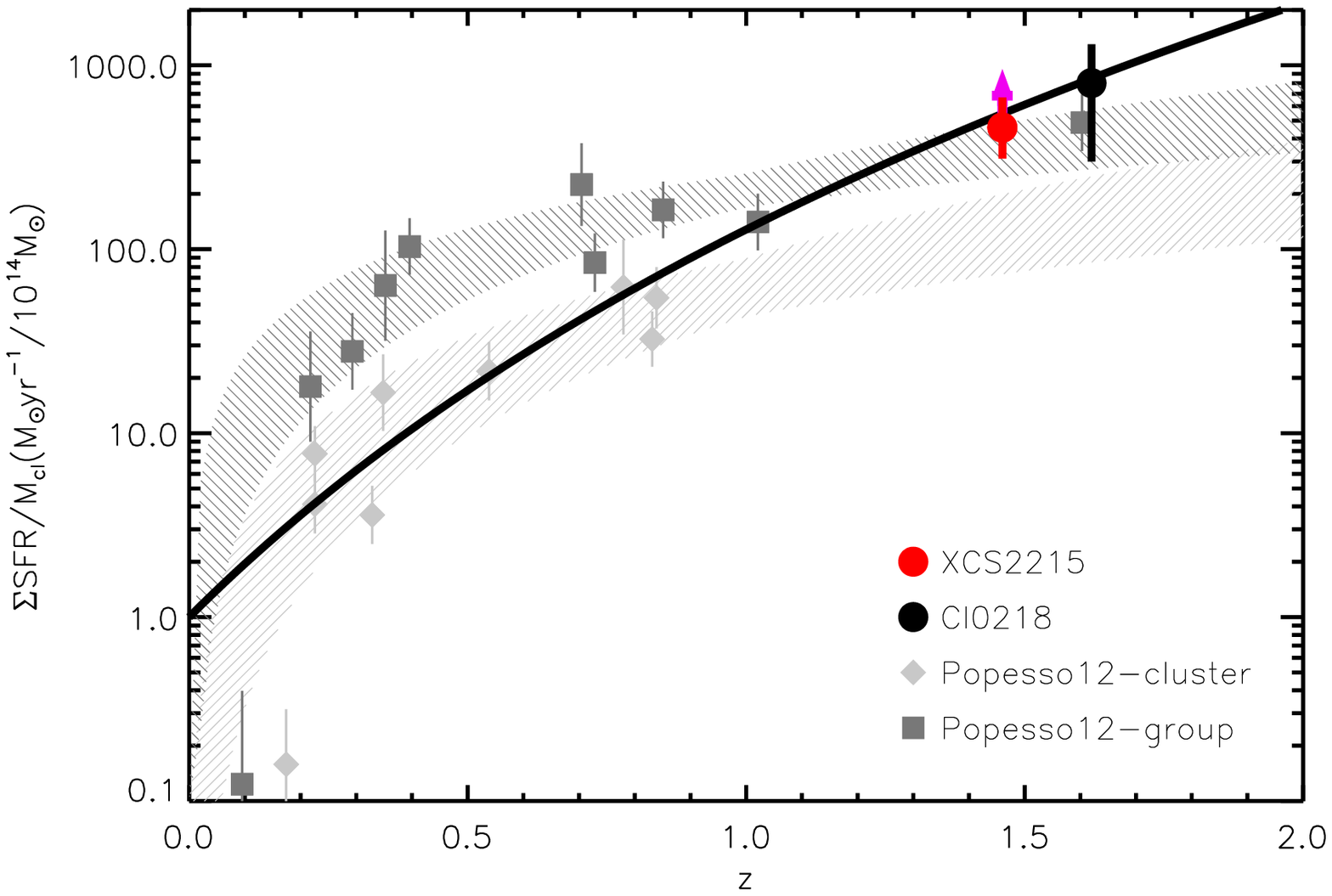}
%\hspace{-1cm}
\caption{The integrated star-formation rate of clusters and groups normalized by their
  total mass. We plot the total SFR of XCS\,J2215
  derived from the 850\,$\mu$m-selected SMGs  within the virial radius of the
  cluster. Including the star formation  in the 24\,$\mu$m sources detected in
  \citet{hilton10} and [O{\sc ii}]\ emitters detected in
  \citet{hayashi10,hayashi14}, the normalized integrated SFR will increase by a
  factor of 1.5 or 2.1 respectively as indicated by the cross and arrow. Compared to the trend of galaxy groups/poor clusters or
  clusters proposed by \citet{popesso11}, the mass-normalized
  SFR of both XCS\,J2215 and Cl\,J0218 \citep[]{smail14} suggest more
star-formation activity in high-redshift clusters,   closer to the power-law
evolution proposed by \citet{geach06}.}\label{fig:sfrz}
\end{figure}

As we have discussed, there appears to be strong evolution in the SFR of galaxies in cluster cores out to high redshift.  To
compare our high-redshift cluster to other low- and intermediate-redshift clusters, we therefore integrate the the total SFR
within the virial radius (0.8\,Mpc) to obtain $\Sigma{\rm SFR}=$\,1400$^{+630}_{-440}$\,M$_{\odot}$\,yr$^{-1}$ and normalize by the cluster mass ($M_{\rm cl}=3\times\,$10$^{14}\,$\,M$_{\odot}$) following \citet{popesso12} to derive a mass-normalized SFR of 460$^{+210}_{-150}\,/\,$10$^{14}$\,yr$^{-1}$ shown in Fig.~\ref{fig:sfrz}. In \citet{popesso12}, the luminosity limit ($L_{\rm IR}\geq $\,10$^{11}$\,L$_{\odot}$ or equivalently SFR\,$\geq$\,20\,M$_{\odot}$\,yr$^{-1}$) of the PACS-detected member galaxies are lower than the luminosity limit of the 850\,$\mu$m-selected SMGs in XCS\,J2215, ($L_{\rm IR} \geq 7\times10^{11}$\,L$_{\odot}$ or SFR\,$\geq$\,130\,M$_{\odot}$\,yr$^{-1}$). Therefore, we also calculated the mass-normalised integrated SFR including the 24\,$\mu$m sources with the limit of SFR\,$\geq$\,100\,M$_{\odot}$\,yr$^{-1}$) from \citet{hilton10}, and the  [O{\sc ii}]\ emitters with the limit of dust-corrected SFR\,$\geq $\,10\,M$_{\odot}$\,yr$^{-1}$ from \citet{hayashi10}. 
 When including these samples in the plot, the mass
normalized integrated SFR increases by a factor of 1.5
and 2.1 respectively compared to that determined solely from  the 850\,$\mu$m-selected SMGs  -- leading to a mass-normalised integrated SFR in galaxies with individual SFRs $\geq $\,10\,M$_{\odot}$\,yr$^{-1}$
of  (950\,$\pm$\,320)\,/\,10$^{14}$\,M$_{\odot}$\,yr$^{-1}$.  
 
Recently, some works \citep[e.g.][]{brodwin13,alberts14} have arguably suggested that $z\sim$\,1.4 is a critical era for star-formation activity in clusters. At $z>1.4$, the clusters selected from the IRACS Shallow Cluster Survey are dominated by active star-forming galaxies, while at $z<1.4$ the star-formation activity in clusters is exceeded by the galaxies in less dense environments. Many studies of individual high-redshift cluster \citep[e.g.][]{bayliss14,smail14,santos14} also show enhanced star-formation activities in clusters at $z>1.4$. \citet{smail14} \citep[see also][]{santos14}
estimated the mass-normalized star-formation density ($\Sigma{\rm SFR}\,/\,M_{\rm  cl}$) of a $z=$\,1.6
cluster, measuring (800\,$\pm$\,400)\,/\,10$^{14}$\,M$_{\odot}$\,yr$^{-1}$,
which is consistent with the total SFR of all cluster
 members in XCS\,J2215: (950\,$\pm$\,320)\,/\,10$^{14}$\,M$_{\odot}$\,yr$^{-1}$.
Both of these clusters lie a factor $\sim$\,4\,$\times$ above the
proposed rich-cluster sequence from \citet{popesso12}, and a factor
$\sim$\,2\,$\times$ above the proposed poor-cluster/group-scaling relation, suggesting that there is more star-formation activity occurring in these high-redshift systems than predicted by extrapolation of these models from lower redshifts.
Indeed, the mass normalized SFRs of both of these
high-redshift clusters is in better agreement with the  simple $(1+z)^7$
evolution proposed by \citet{geach06} \citep[see also][]{cowie04}.

However, as \citet{geach06} point out, the factor $\gsim$\,2
scatter in mass-normalized SFR  between the clusters at a fixed redshift suggests that
individual cluster environments have strong influences on the star-formation histories of their constituent galaxies.  In particular, the enhanced star-formation activities may be caused by the merging of less massive clusters or groups \citep[as seen in the merging system at low redshift e.g.][]{marcillac07}, and do not happen in the more relaxed systems \citep[e.g. a cluster at slightly lower redshift at $z=1.39$ in][]{bauer11}. At present we have too few high-redshift clusters to determine if there is similarly strong variation in levels of cluster activity as seen at lower redshifts, or if all high-redshift clusters display similarly levels of activity.  Further deep sub-millimeter observations of $z\geq $\,1 clusters are required to test this.
However, there is already some evidence for differences between the two $z\sim$\,1.5 clusters studied with SCUBA-2:  the majority of the SMGs discovered in XCS\,J2215 lie within the core region of the cluster interspersed with the passive cluster population, whereas in Cl\,J0218 \citet{smail14} found that the cluster core was dominated by passive galaxies, with the SMGs lying on the outskirts.  There are also potential differences in the apparent masses of the SMG populations in these
two clusters, with \citet{smail14} finding a median restframe $H$-band absolute magnitude (which is a crude proxy for stellar mass) for their 
850\,$\mu$m-detected SMGs of $M_H\sim -$24.0, 
compared to $M_H=-$25.0\,$\pm$\,0.3 for the SMGs in XCS\,J2215.  The XCS\,J2215 SMGs are therefore brighter (and potentially more massive?) than those in Cl\,J0218, and so they ought to
evolve into more luminous galaxies at the present day, assuming similar evolutionary pathways.  Indeed, adopting a canonical burst lifetime of 100\,Myrs \citep[e.g.][]{simpson14}, we would
expect these galaxies to fade by 3.2 magnitudes in the restframe $H$-band by the present day,  assuming no subsequent star formation, merging or stripping.  The
median $H$-band absolute magnitude of the XCS\,J2215 SMGs would then correspond to $M_H\sim -$21.8 -- the brightness of a typical elliptical galaxy at $z\sim$\,0.  Along with their central location, this perhaps suggests that we are seeing the final stages of the formation of the bulk of the elliptical galaxy population in XCS\,J2215, whereas Cl\,J0218, with its passive-dominated core, has already passed through this phase.

%
%fig 6
%

\section{Conclusions}
\label{sec:conclusions}

We have obtained deep 450- and 850-$\mu$m imaging of the $z=$\,1.46
X-ray selected cluster XCS\,J2215.9$-$1738 using the SCUBA-2 bolometer
camera on the JCMT.  We combine these new observations with multiwavelength archival data on this system, including {\it Spitzer} MIPS mid-infrared imaging, {\it Herschel} PACS far-infrared data and VLA radio observations.  Together these data provide a sensitive survey of ULIRG
activity in the central regions of the cluster.

From these data we identify a significant over-density of eleven SMGs
in the cluster core.  By both statistically-matching these sub-millimeter sources to
available MIPS 24\,$\mu$m and 1.4\,GHz radio sources and deblending the
SCUBA-2 maps using the MIPS and radio catalogs as priors,  we determine
robust  identifications of counterparts to all eleven
SMGs (the majority from radio identifications).

Two of these SMG counterparts have precise spectroscopic redshifts
which confirm that they are members of the cluster.  Three others are
identified as likely cluster members through their detection in
narrow-band imaging of the redshifted [O{\sc ii}] emission line, while
a further three have photometric redshifts which are consistent with
their being members.  Hence in total eight of the eleven SMG
counterparts, including all of those within the virial radius (0.8\,Mpc),
are probable or potential cluster members.  

We fit the far-infrared and radio SEDs of the cluster members and
derive a typical luminosities $L_{\rm IR}\sim (1.0\pm0.1)\times10^{12}$\,L$_\odot$ (corresponding
to a SFR\,=\,170\,$\pm$\,20\,M$_{\odot}$\,yr$^{-1}$).  These cluster SMGs
appear to have temperatures which are $\sim$\,3--4\,K cooler than
comparably luminous low-redshift ULIRGs, perhaps indicating a more extended
distribution of dust in SMGs compared to local ULIRGs.  We also show that the galaxies lie within the scatter of the radio--far-infrared correlation seen for field SMGs at high redshifts.

Finally, to investigate the evolution of the starburst population in
the cores of clusters with redshift, we integrate the star formation
within all galaxies in the cluster core (through our 850$\mu$m-selected SMGs, 24$\mu$m-selected galaxies and [O{\sc ii}] emitters).
Normalizing by the total cluster mass, we show that XCS\,J2215 contains
one of the most active clusters cores studied to date.  By combining
with other low- and high-redshift samples, we show that the evolution
of the mass-normalized SFR appears to continue to 
increase at $z\geq $\,1, consistent with an evolutionary trend scaling as
$(1+z)^n$ with $n\sim $\,7.  However, we  have observations of very few clusters at $z>$\,1 and so it is unclear
if the scatter between clusters that is seen in the
low-redshift samples at fixed redshift (reflecting a variety of evolutionary
states),  also exists at
high redshift. The evidence we have from the differences in the environments and near-infrared
luminosities of the SMGs in the two $z\sim$\,1.5 clusters studied 
with SCUBA-2 hints that evolutionary differences may already exist between clusters
at this epoch, reflecting their different levels of development.

\section*{Acknowledgments}\label{lastpage}

CJM acknowledges support from ERC Advanced Investigator programme
DUSTYGAL 321334.  IRS acknowledges support from STFC (ST/L00075X/1),
the DUSTYGAL ERC programme and a Royal Society/Wolfson Merit
Award. AMS acknowledges an STFC Advanced Fellowship through grant
number ST/H005234/1 and the Leverhume Trust.  The James Clerk
Maxwell Telescope is operated by the Joint Astronomy Centre on behalf
of the Science and Technology Facilities Council of the United
Kingdom, the National Research Council of Canada, and (until 31 March
2013) the Netherlands Organisation for Scientific Research. Additional
funds for the construction of SCUBA-2 were provided by the Canada
Foundation for Innovation.  This work is based [in part]
on observations made with the {\it Spitzer Space Telescope}, which is
operated by the Jet Propulsion Laboratory, California Institute of
Technology under a contract with NASA. The National Radio Astronomy
Observatory is a facility of the National Science Foundation operated
under cooperative agreement by Associated Universities, Inc.  This
research has made use of NASA's Astrophysics Data System.

%
%table 1
%
%\clearpage
\begin{deluxetable*}{lccccccccccc}
\tabletypesize{\scriptsize}
%\rotate 
\tablewidth{0pc}
\tablecolumns{12} 
\tablecaption{Properties of the 850\,$\mu$m sources within 1\,Mpc radius of  XCS\,J2215.\tablenotemark{*} \label{table:850srcs}}
\tablehead{ 
\colhead{ID}   & \colhead{R.A.\ }         & \colhead{Dec.\ }      &\colhead{$f_{850}$} & \colhead{$f_{450}$} & \colhead{$f_{160}$} & \colhead{$f_{70}$} & \colhead{$f_{24}$} & \colhead{$f_{1.4}$} & \colhead{$r_{\rm c}$}  & \colhead{$P_{24}$} & \colhead{$P_{1.4}$} \\
       & \colhead{(J2000)} & \colhead{(J2000)} &   \colhead{(mJy)}      & \colhead{(mJy)}         & \colhead{(mJy)}        & \colhead{(mJy)}       & \colhead{(mJy)}      & \colhead{($\mu$Jy)}             & \colhead{(Mpc)}           &                   &  
}
\startdata
%     	     &RAD           &DECD          &f_{850}           &s4flx                    &PACSRflx                &PACSBflx       &mips24flx      &S1400       &distMPC
04   &22\,15\,58.4&$-$17\,38\,19& $3.2 \pm0.6$ & $<16           $ & $8.1   \pm1.4 $  & $1.0 \pm0.4$ & $0.17\pm0.01$ & $33\pm8$ & 0.15   & 0.13   & 0.05 \\
06   &22\,15\,59.8&$-$17\,37\,59& $2.9 \pm0.6$ & $16 \pm5$ & $8.6   \pm1.4 $  & $1.3 \pm0.4$ & $0.22\pm0.01$ & $76\pm8$ & 0.07     & 0.03 & 0.01 \\
11 &22\,15\,58.7&$-$17\,37\,47& $2.2 \pm0.6$ & $18 \pm5$ & $9  \pm4  $  & $1.5 \pm0.5$ & $0.09\pm0.01$ & $86\pm8$ & 0.16            & 0.02 & 0.01 \\
13 &22\,15\,57.3&$-$17\,37\,55& $1.9 \pm0.6$ & $16 \pm5$ & $10.0 \pm1.4$   & $2.2 \pm0.4$ & $0.22\pm0.01$ & $52\pm9$ & 0.24     & 0.03 & 0.01 \\
\noalign {\smallskip} 
02  &22\,16\,02.9&$-$17\,38\,35& $5.9\pm0.7$   & $21\pm6$  & $19.4\pm1.4$ & $7.8  \pm0.4$ & $0.72\pm0.01$ & $63\pm9$ & 0.55        & 0.05 & 0.02  \\
03  &22\,16\,00.6&$-$17\,38\,35& $3.4 \pm0.7$  & $17\pm6$   & $7.5  \pm1.4$ & $1.7  \pm0.4$ & $0.23\pm0.01$ & $66\pm9$ & 0.32       & 0.07 & 0.04 \\
05  &22\,15\,59.8&$-$17\,37\,19& $3.1\pm0.7$   & $<17     $     & $<4             $ & $<1.3       $      & $0.20\pm0.01$ & $<25$      & 0.39           & 0.01  & ...   \\
10&22\,16\,04.8&$-$17\,37\,51&$3.0\pm0.8$    & $24\pm6$  & $10.6\pm1.3$ & $2.2  \pm0.5$ & $0.28\pm0.01$ & $53\pm8$ & 0.69        & 0.04  & 0.01 \\
\noalign {\smallskip} 
01   &22\,15\,59.0&$-$17\,39\,43& $6.6 \pm0.8$  & $26\pm6$ & $10.5 \pm1.5$ & $2.8  \pm0.4$ & $0.56\pm0.01$ & $168\pm10$ & 0.84   & 0.01    & 0.003\\
07   &22\,16\,01.2&$-$17\,39\,35& $3.5\pm0.8$   & $21\pm7$ & $10.0 \pm1.4$ & $1.6  \pm0.5$ & $0.46\pm0.01$ & $47\pm10$ & 0.82      & 0.000  &0.000 \\
12 &22\,16\,02.0&$-$17\,39\,47& $2.7\pm0.8$   & $22\pm7$ & $8.4   \pm1.5 $& $3.1  \pm0.4$ & $1.30\pm0.01$ & $<29     $ & 0.98            & 0.05   & ...

\enddata
  \tablenotetext{*}{\small The columns lists the ID, the 850\,$\mu$m position and fluxes  at 850\,$\mu$m and 450\,$\mu$m measured using SCUBA-2, with PACS on {\it Herschel} at 160\,$\mu$m and 70\,$\mu$m, MIPS on {\it Spitzer} at 24\,$\mu$m, and JVLA at 1.4\,GHz.  For  non-detections with S/N\,$<$\,3, the 3-$\sigma$ upper limit is listed.  The next column gives the cluster-centric radius of each source in Mpc.  The 850\,$\mu$m sources all have close and bright 1.4\,GHz and/or 24\,$\mu$m counterparts and the final two columns are the $P$-statistics of these counterpart at 24\,$\mu$m and 1.4\,GHz.  The sources are grouped in the table according to the projected distance from the cluster core (r$_{\rm c}$). Within each group, the sources are ranked according to their 850\,$\mu$m fluxes.}  
\end{deluxetable*}
  
%
%table 2
%
\begin{deluxetable*}{lccccc|ccc|ccl}
\tabletypesize{\scriptsize}
%\rotate
\tablewidth{0pc}
\tablecolumns{12} 
\tablecaption{Derived properties for the 850\,$\mu$m sources in XCS\,J2215.\tablenotemark{*} \label{table:sfr}}
\tablehead{ 
\colhead{ID}   & \colhead{R.A.\ }         & \colhead{Dec.\ } & \colhead{$z$}    & \colhead{$L_{\rm IR}$}        & \colhead{$T_{\rm d}$}       &    \multicolumn{3}{c}{SFR}  &  \colhead{m$_{\rm 4.5\mu m}$} &  \colhead{$(z'-K)$}         & \colhead{Notes}         \\
\colhead{}       & \colhead{}                    & \colhead{}           & \colhead{}          & \colhead{}                               & \colhead{}                             & \colhead{$_{\rm FIR}$} &\colhead{$_{\rm [OII]}$}   &\colhead{$_{\rm 24{\mu}m}$}  & \colhead{}                             & \colhead{}                             & \colhead{}                             \\
       &  \colhead{(J2000)} & \colhead{(J2000)} &    & \colhead{(10$^{11}$L$_\odot$)}  &  \colhead{(K)}             &\multicolumn{3}{c}{(M$_\odot$\,yr$^{-1}$)}   &  \colhead{(AB)}     & \colhead{(AB)} &   
}
\startdata
04      & 22\,15\,58.10 & $-$17\,38\,14.1 & [O{\sc ii}] & $7.5_{-2.5}^{+1.6}$ & $28\pm1$ [23]   & $130_{-40}^{+30}$ & 8    &  170   & 19.9  & 2.38  &     \\      
06      & 22\,15\,59.64 & $-$17\,37\,59.0 & 1.469& $8.0_{-1.7}^{+1.4}$ & $29\pm1$ [23]   & $140_{-30}^{+20}$        & 8   & 240 & 19.1  & 2.04    & H10: 744/747.    \\
11     & 22\,15\,58.44 & $-$17\,37\,46.6 & [O{\sc ii}] & $8.3_{-2.9}^{+2.1}$ & $31\pm3$ [24]   & $140_{-50}^{+40}$   & 11 &  110 & 19.7  & 2.51  &   \\
13     & 22\,15\,57.17 & $-$17\,37\,52.9 & 1.454& $8.6_{-1.5}^{+2.1}$ & $32\pm2$ [24]   & $160_{-30}^{+30}$         & 11  & 160 & 20.3 & 1.76  &   H10: 35. \\
\noalign {\smallskip} 
02      & 22\,16\,03.07 & $-$17\,38\,40.0 & phot-$z$  & $21.2_{-2.0}^{+1.5}$ & $29\pm1$ [27]   & $360_{-50}^{+30}$ & ...   & 650  & 18.8 & 1.94  &  H10: 1022.      \\
03      & 22\,16\,00.87 & $-$17\,38\,30.8 & [O{\sc ii}] & $8.3_{-1.7}^{+1.7}$ & $28\pm1$ [22]   & $130_{-30}^{+30}$     & 10 & 160  & 20.4 & 1.99    &   H10: 983.  \\
05      & 22\,15\,59.75 & $-$17\,37\,17.2 & phot-$z$  & $10.6_{-5.8}^{+4.1}$ & $26\pm2$ [23]   & $150_{-80}^{+250}$ & ... & 140  & 21.1 & 0.92  &    only 850$\mu$m in FIR \\
10     & 22\,16\,04.92 & $-$17\,37\,54.1 & phot-$z$   & $10.2_{-2.0}^{+1.7}$ & $29\pm1$ [24]   & $180_{-150}^{+200}$& ... & 220 & 20.0 & 2.02  &  H10: 709.     \\
\noalign {\smallskip} 
01      & 22\,15\,58.95 & $-$17\,39\,42.3 & ...   & $10.9_{-1.5}^{+3.0}$ & $25\pm1$ [21]   & $180_{-20}^{+80}$                 & ...  & 510 & ... & ...  &  Radio-bright AGN.  \\
07      & 22\,16\,01.13 & $-$17\,39\,34.9 & ...   & $9.4_{-1.6}^{+1.9}$   & $29\pm2$ [23]   & $160_{-30}^{+30}$                 & ...   & 380 & ... & ...    &   X-ray AGN.  \\
12    & 22\,16\,02.27 & $-$17\,39\,50.1 & ...   & $9.6_{-1.9}^{+2.1}$   & $29\pm2$ [23]   & $160_{-30}^{+30}$                   & ...   & 1200 & ... & ...     &  X-ray AGN.     

\enddata
\tablenotetext{*}{\small The columns lists the ID, the F814W position of the proposed counterpart, the redshift or source of the membership, the far-infrared luminosity and dust temperature, the star-formation rate, the  4.5\,$\mu$m IRAC magnitude and $(z'-K)$ color and finally notes on each candidate SMG.  In addition to the SFR calculated using the far-infrared SED in this paper, the two SFRs calculated using the flux of [O{\sc ii}] \citep{hayashi10,hayashi14}, and 24$\mu$m are listed for comparison. The SFR Counterparts in \citet{hilton10} are indicated by H10 and their IDs. In the column listing $T_{\rm d}$, the second value (in parenthesis) corresponds to the dust temperature derived by taking the wavelength of the peak of the best-fitting dust SED template and assuming $\lambda_{\rm peak}T_{\rm d}= $\,2.897\,$\times$\,10$^3$\,m~K.}
\end{deluxetable*}

\end{document}